\documentclass[twocolumn,showpacs,preprintnumbers,amsmath,amssymb]{revtex4}
\usepackage{epsfig}
\usepackage{dcolumn}
\usepackage{bm}

\makeatletter

 \def\Tnl{t^{'}_{\rm nl}}
 \def\tnl{t_{\rm nl}}
 \def\Ki{k_{\rm i}}
 \def\Ku{k_{\rm u}}
 \def\Ks{k_{\rm s}}

 \def\bb{\begin{equation}}
 \def\ee{\end{equation}}
 \def\bba{\begin{eqnarray}}
 \def\eea{\end{eqnarray}}
 
 \def\rf(#1){(\ref{#1})}
 \long\def\hide"#1"{}
 \def\ms1{{\kern .035em s}}

\begin{document}

\preprint{DECAYPAPER-REVTEX-TEST-1}
\title{The global picture of self-similar and not self-similar
decay in Burgers Turbulence}

\author{Alain Noullez}
 \email{anz@obs-nice.fr}
 \affiliation{Observatoire de la C\^ote d'Azur, Lab. Cassiop\'ee,\\
B.P.~4229, F-06304 Nice Cedex 4, France
}

\author{Sergey~N. Gurbatov}
 \email{gurb@rf.unn.ru}
 \altaffiliation[Also at ]{Observatoire de la C\^ote d'Azur, Lab. Cassiop\'ee,
B.P.~4229, F-06304 Nice Cedex 4, France}
\affiliation{Radiophysics Dept., University of Nizhny Novgorod,\\
23, Gagarin Ave., Nizhny Novgorod 603950, Russia
}

\author{Erik Aurell}
 \email{erik.aurell@physics.kth.se}
 \affiliation{Dept. of Physics, KTH - Royal Institute of Technology,\\
AlbaNova University Center, SE-106 91 Stockholm, Sweden
}

\author{Sergey~I. Simdyankin}
\affiliation{Department of Chemistry, University of Cambridge,\\
Lensfield Road, Cambridge CB2 1EW, UK}
\altaffiliation[Also at ]{Radiophysics Dept., University of Nizhny Novgorod,\\
23, Gagarin Ave., Nizhny Novgorod 603950, Russia}

\date{\today}


\begin{abstract}

This paper continue earlier investigations on the decay of Burgers turbulence
in one dimension from Gaussian random initial conditions of the power-law
spectral type~$E_0(k)\sim|k|^n$.  Depending on the power $n$, different
characteristic regions are distinguished.  The main focus of this paper is to
delineate the regions in wave-number $k$ and time $t$ in which self-similarity
can (and cannot) be observed, taking into account small-$k$ and large-$k$
cutoffs.  The evolution of the spectrum can  be inferred using physical
arguments describing the competition between the initial spectrum and the new
frequencies generated by the dynamics.  For large wavenumbers, we always have
$k^{-2}$ region, associated to the shocks.  When~$n$ is less than one, the
large-scale part of the spectrum is preserved in time and the global evolution
is self-similar, so that scaling arguments perfectly predict the behavior in
time of the energy and of the integral scale.  If~$n$ is larger than two, the
spectrum tends for long times to a universal scaling form independent of the
initial conditions, with universal behavior $k^2$  at small wavenumbers. In the
interval $2<n$ the leading behaviour is self-similar, independent of $n$ and 
with universal behavior $k^2$  at small wavenumber.  When $1<n<2$, the spectrum
has three scaling regions\,: first, a $|k|^n$ region at very small $k$\ms1 with
a time-independent constant, second, a $k^2$ region at intermediate
wavenumbers, finally, the usual $k^{-2}$ region.  In the remaining interval,
$n<-3$ the small-$k$ cutoff dominates, and $n$ also plays no role.  We find
also (numerically) the subleading term $\sim k^2$ in the evolution of the
spectrum in the interval $-3<n<1$. High-resolution numerical simulations have
been performed confirming both scaling predictions and analytical asymptotic
theory.

\end{abstract}
\maketitle

\section{Introduction}
\label{s:introduction}

We study here Burgers equation 
\begin{equation}
\frac{\partial v}{\partial t} + v\,\frac{\partial v}{\partial x} = \nu
\frac{\partial^2 v}{\partial x^2}
\label{BE}
\end{equation}
in the limit of vanishing coefficient $\nu$.
First introduced 
by J.M.~Burgers as a model
of hydrodynamic turbulence, this equation
arises in many situations 
in physics, see \cite{Burgers,Whitham,RudenkoSoluyan,Kuramoto,Gurbatov,WW98} 
for classical (and more recent) monographs. 
It is fair to say that one  of the  main interest in Burgers equation
over the last decade has been
as a model for structure formation in the early
universe within the so-called adhesion 
approximation \cite{GurbatovSaichev,Shandarin,Vergassola}. 
The Hopf-Cole transformation, to which we will
return below, has been developped into a powerful tool
to elucidate the statistical properties of solutions
to Burgers equations with random initial 
conditions of cosmological type \cite{Sinai,SheAurellFrisch,Vergassola}.
If a random force is added to the right-hand side
of (\ref{BE}), the resulting KPZ equation is
one of the most important models of e.g.
surface growth \cite{Schwartz,Kardar,BS95}. 

Investigations of Burgers turbulence have a long pre-history, started already
by Burgers (1974), who was mainly concerned with white noise initial
conditions.  But nevertheless only recently \cite{Fra00}
the exact statistical properties of the Burgers equation for the case $n=0$
were found.  The case of 
fractal Brownian motion for the potential or for initial velocity 
is much more complicated  \cite{Molchan}.  

Burgers equation (\ref{BE}) describes two principal
effects inherent in  any turbulence \cite{Frisch}: the nonlinear
redistribution of energy over the spectrum and the
action of viscosity in  small-scale regions.
Except for the direct physical application, Burgers
equation is hence also of great interest to test theories
and models of fully developped turbulence. 
This paper follows that tradition. 
In an earlier
contribution \cite{GSAFT97} we showed how self-similarity
arguments, going back to Kolmogorov \cite{Kolmogorov1941b}
and Loitsyanski \cite{Loitsyanski} can be be disproved,
in Burgers equation, for a class of initial conditions.
A similar result was later arrived at by Eyink and 
Thompson for the Navier-Stokes equation
\cite{Eyink00}, within an eddy-damped, quasi-normal
Markovian (EDQNM) scheme.
In this paper, we will discuss in greater detail how the self-similar
(and not self-similar) regimes are realized with initial
conditions that are only self-similar over a finite range.
The range in which self-similarity can be observed
(or not observed) changes in wave-number space with time,
in a way that depends both on the initial spectral slope,
and on the low-$k$ and high-$k$ cutoffs in the initial
data.
 
The paper is organized as follows\,: in section~\ref{s:large-time decay} we
recall the basic properties of Burgers equation, and give a more precise
description of the class of initial conditions we consider.  In
section~\ref{s:phivel} ($n>1$ and $n<-3$) we consider the situation when both
the velocity and the velocity potential are homogeneous  Gaussian processes. 
For such initial conditions, we have asymptotically self-similar evolution with
universal behavior of the spectrum $E(k)\sim k^2$ and $E(k)\sim k^{-2}$ at
small and large wavenumber respectively.  For $1<n<2$, the spectrum, at long,
but finite time, has also the region $|k|^n$ at very small $k$ with
time-independent constant, but followed by a region~$k^2$ which quickly becomes
dominant.  In section~\ref{s:psi} ($-1<n<1$) we consider the case of
homogeneous velocity potential.  In section~\ref{s:nonpsi} ($-3<n<-1$) we
consider the case of non homogeneous velocity potential.  In the last two cases
the long-time evolution of the spectrum is self-similar in some region of
($k,t$) plane even we have cutoff wavenumber at small and large wavenumber.  In
section~\ref{s:discussion} we summarize and discuss our results.  Details of
the numerical methods are presented in appendix~\ref{s:numerical}.

\section{Large-time decay, self-similarity and Burgers phenomenology}
\label{s:large-time decay}

We study in this paper the evolution of the velocity field, when the initial
conditions are random and the initial power spectral density  is self-similar,
that is of the form of a power-law~$E_0(k)\sim|k|^n$ .  Let us suppose this is
the case for a finite interval $\Ki\le |k| \le \Ku$, where $\Ki$ and $\Ku$  are
cutoff wavenumbers at large and small  scales respectively - on the infrared
and ultraviolet part of the energy spectrum.  We assume the spectrum to go to
zero faster than any power-law on either side.  We are then interested in the
plane $(k,t)$, and specifically in the following question: where is the
behaviour ``universal'', that is, explainable in terms of a few global
quantities, and where will the specific values of $n$, $\Ki$ and $\Ku$ play an
essential role~?

\begin{figure*}
 \centerline{
 \epsfig{file=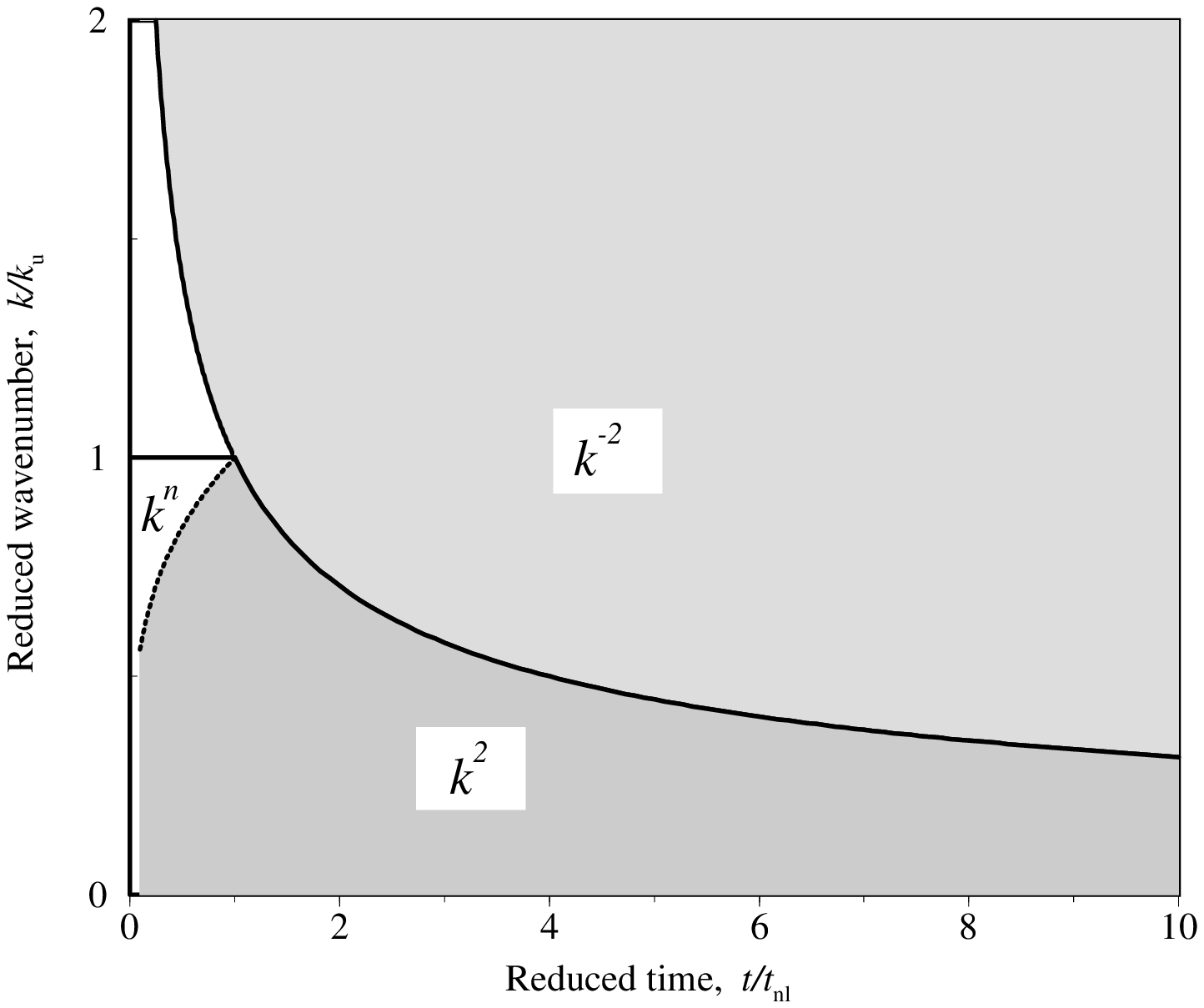,width=6cm}\hspace{3mm}(a)\hspace{12mm}
 \epsfig{file=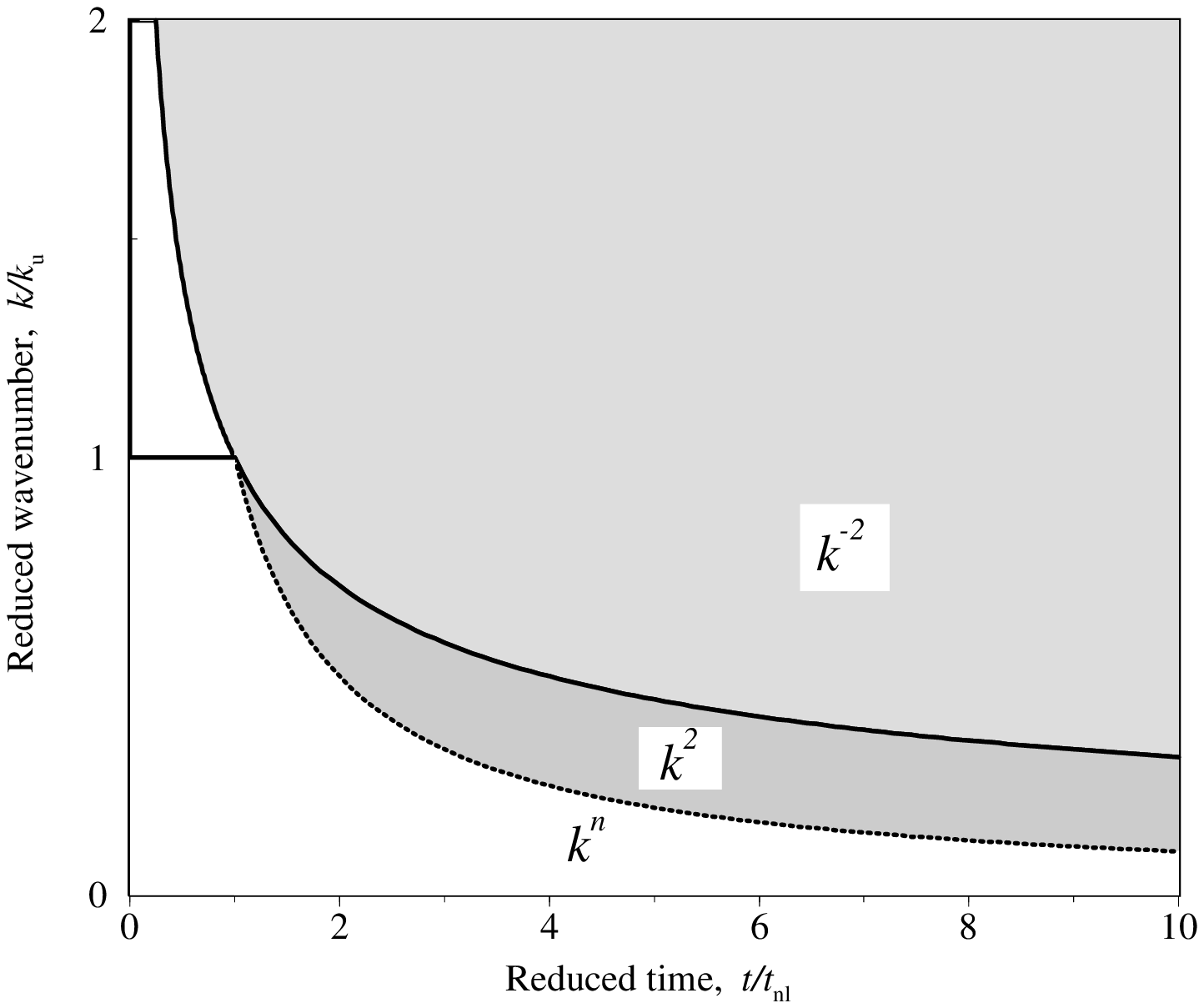,width=6cm}\hspace{3mm}(b)}
 \vspace{8mm}
 \centerline{
 \epsfig{file=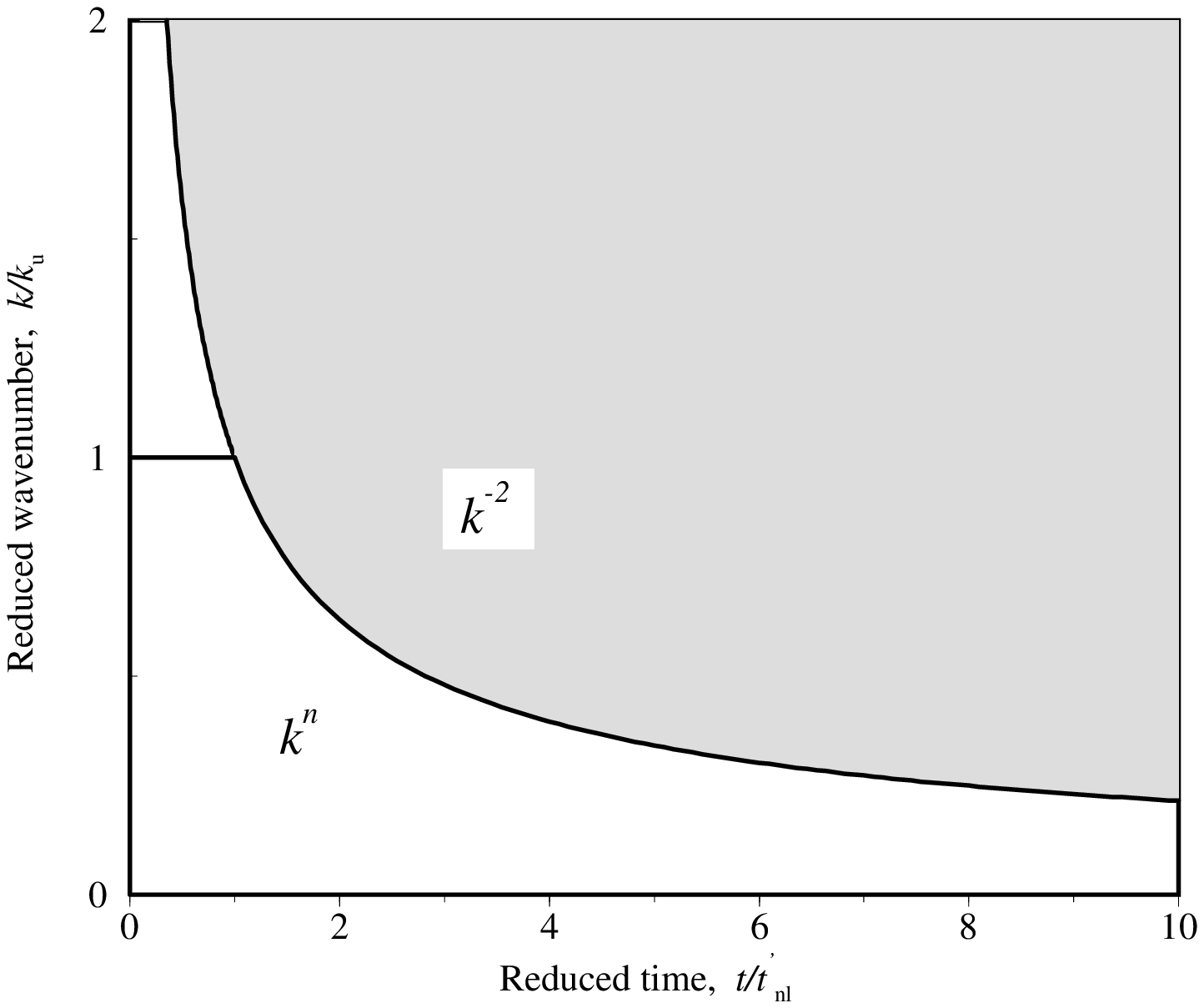,width=6cm}\hspace{3mm}(c)\hspace{12mm}
 \epsfig{file=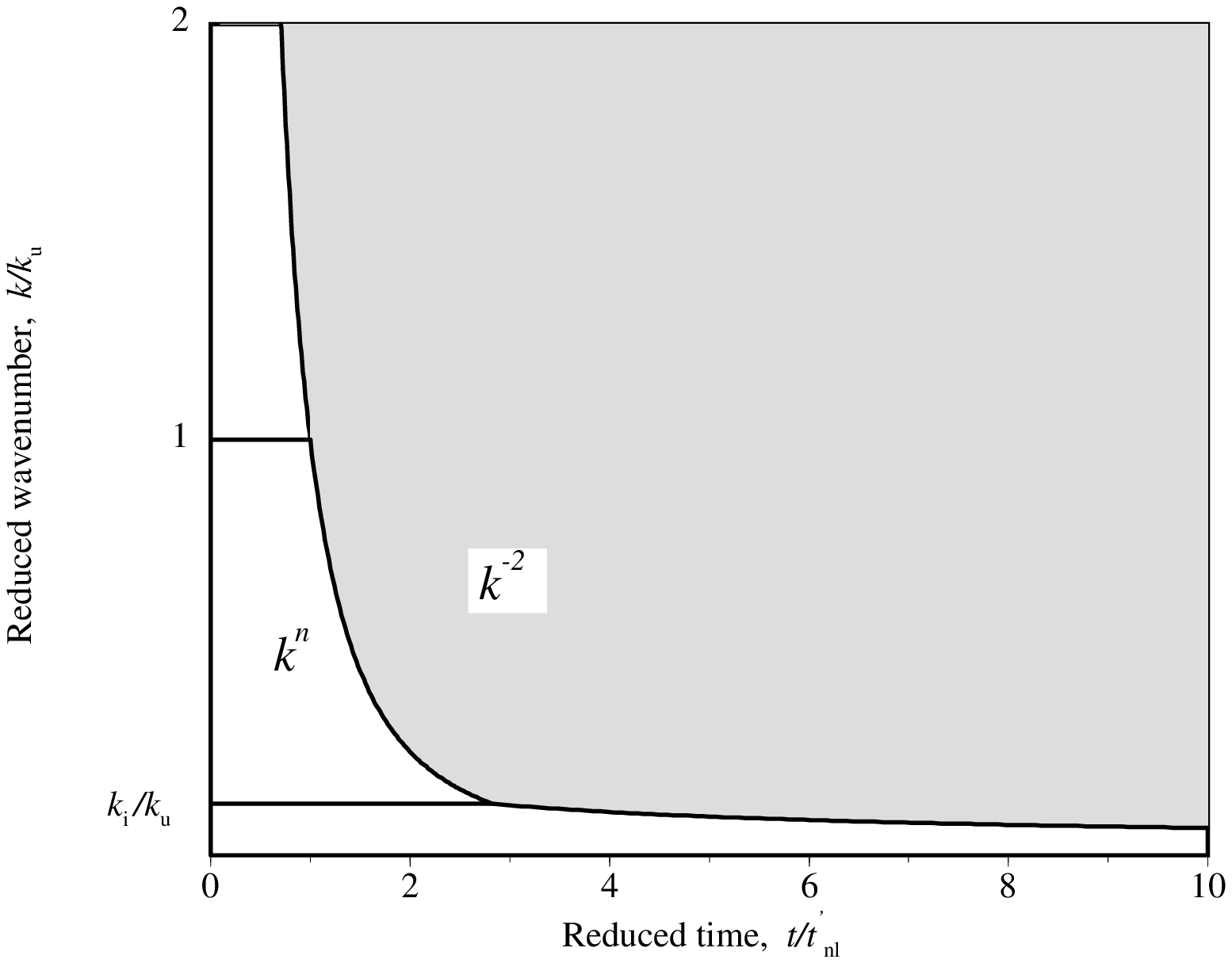,width=6cm}\hspace{3mm}(d)}
 \vspace{6mm}
 \caption{The universal behavior $E(k,t)\sim |k|^n, k^2, k^{-2}$ and
self-similarity $E(k,t) = L^3(t)t^{-2} \tilde E(kL(t))$ of the energy 
spectrum on the $(k-t)$ plane. The line $k_L(t) \sim 1/L(t)$ is the border
between $k^{-2}$ and $|k|^n$ or $k^2$ asymptotics of the spectrum. 
The line $\Ks(t)$ is  the border
between $|k|^n$ and $k^2$ behavior at smal wavenumber. 
Fig.~\ref{fig:summary}~a - power index $n>2$\,: universal self-similar behavior 
at $t \gg \tnl$.  Fig.~\ref{fig:summary}~b power index $1<n<2$\,: 
self-similar behavior only when $t \gg \tnl$ and $k \gg \Ks(t)$. 
Fig.~\ref{fig:summary}~c power index  $-3<n<1$\,: self-similar behavior 
for $t \gg \tnl$.
Fig.~\ref{fig:summary}~d power index  $-3<n<1$ and an infrared cutoff~$\Ki$\,:
self-similar behavior for $t_{\rm i} \gg t \gg \tnl$ and $\Ki \gg k \gg \Ku$.}
 \label{fig:summary}
\end{figure*}

Fig.~\ref{fig:summary} illustrates the results we will
show. 

Let us now proceed to explain how Fig.~\ref{fig:summary} can be motivated. From
equation (\ref{BE}) we can derive an equation for the velocity potential $\psi$
and by the Hopf-Cole transformation $\psi = 2 \nu \ln \theta$ \cite{Hopf,Cole},
we turn this into a linear diffusion equation for an
auxiliary field $\theta$.
Convolution of the
initial data for $\theta$ with 
the standard heat kernel gives the
the solution of diffusion 
equation,
which in the limit $\nu \rightarrow 0$ may be 
computed by the method of steepest descent.  
The velocity potential in the limit 
$\nu \rightarrow 0$ is then 
\begin{equation}
\psi(x,t) = \max_a \left[ \psi_0(a) - \frac{(x-a)^2}{2t}\right].
\label{MAX}
\end{equation}
where $\psi_0(a)$ is the initial potential. The velocity
field follows by differentiation in $x$ and reads
\begin{equation}
v (x,t) =  - \psi_x(x,t) =  \frac{ x-a(x,t) } {t} ,
\label{velocity111}
\end{equation}
where $ a(x,t) $ is the argument of the maximization
in (\ref{MAX}) for given given $x$ and $t$. If there are
several such  $a(x,t)$'s, we are at a shock, where the
velocity field is discontuous.  See e.g.\cite{Burgers,Whitham,Gurbatov,WW98}
or {\it op. cit.} for an in-depth discussion.

In this paper we look at initial conditions with energy spectrum
\begin{equation}
E_0(k) = \alpha^2 |k|^n b_0(k),
\label{initsp}
\end{equation}
where $n$ is the spectral exponent and 
$b_0(k)$ satisfies $b_0(k) = 1$ in some regions of wavenumbers 
[$\Ki,\Ku$], 
while going quickly to zero on either side.
The (mean) energy is
\begin{equation}
E(t)\equiv \langle v^2(x,t)\rangle = \int_{-\infty}^\infty E(k,t)\,dk.
\label{defet}
\end{equation}
and the initial energy is denoted
\begin{equation}
\sigma_v^2 \equiv \langle v_0^2(x)\rangle = \int_{-\infty}^\infty E_0(k)\,dk.
\label{defsigmav}
\end{equation}

It is clear from formula (\ref{MAX}) above
that the solutions depend solely from the initial velocity potential.
Let us introduce the variance of the initial potential (if it exists)\,:
\begin{equation}
\sigma_{\psi}^2 \equiv \langle \psi_0^2(x)\rangle =
\int_{-\infty}^\infty {E_0(k)\over k^2}\,dk.
\label{defsigmapsi}
\end{equation}
It is also clear that for a 
continuous initial  velocity field the time of first shock
formation  depends on the initial velocity
gradients as
$t_{\rm s,first}=-1/\min\partial_x v(x,t_0)$. Consequently,
the variance of the initial velocity gradient (if it exists)
\begin{equation}
\sigma_{q}^2 \equiv \langle (\partial_x v_0(x))^2 \rangle =
\int_{-\infty}^\infty k^2 E_0(k)\,dk.
\label{defsigmau}
\end{equation}
should also be of interest as it determines the typical
time of first shock formation  $t_{\rm s}=1/\sigma_{q}$.
Since the initial conditions are scaling only in a finite range,
all three characteristic quantities $\sigma_{q}$,  $\sigma_v$
and $\sigma_{\psi}$ exist and are finite. This situation we have 
always in  numerical experiments 
when $\Ki$ is determined by the size of box $L_{\rm box}$ and $\Ku$ 
is the inverse of the step of discretization. However, depending 
on~$n$, they are dominated by one or the other
of the the cutoffs. This suggests
hence the following first division of spectral exponents $n$
(see Table~\ref{t:n-division}).

\begin{table}
\begin{tabular}{l|cccccccc}\hline
$n$ & \multicolumn{3}{r}{\qquad -3 \qquad} &
\multicolumn{2}{r}{\qquad -1} &
\multicolumn{2}{r}{\qquad 1} & \qquad \\ \hline 
$\sigma_{q}$    && $\Ki$ && $\Ku$ && $\Ku$ && $\Ku$ \\ \hline 
$\sigma_v$      && $\Ki$ && $\Ki$ && $\Ku$ && $\Ku$ \\ \hline 
$\sigma_{\psi}$ && $\Ki$ && $\Ki$ && $\Ki$ && $\Ku$ \\ \hline 
\end{tabular}
\caption{Division of domains of spectral exponent~$n$ according to second
moments of $\partial_x v_0$, $v_0$ and $\psi_0$. An entry $\Ki$ and $\Ku$
indicates that the corresponding quantity is dominated respectively by
the low-$k$ and high-$k$ cutoffs.}
\label{t:n-division}
\end{table}

 From the maximum representation of the solutions to Burgers equation
(\ref{MAX}), we can introduce the scale $L(t)$, proportional to the typical
value of $|(x-\tilde{a}(x,t)|$. For large time, balancing the two terms
in~(\ref{MAX}), we have the following prediction for the scale $L(t)$ of
Burgers turbulence (see Table~\ref{t:n-Burgers-division}).
\begin{table} 
\begin{tabular}{l|cccccccc}\hline
$n$ &&& \multicolumn{1}{r}{-3} &&
\multicolumn{1}{r}{-1} &&
\multicolumn{1}{r}{1} & \qquad \\ \hline 
$L(t)$ &\quad& $(\sigma_{\psi}t)^{1/2}$ &\qquad& $(\alpha t)^{2/(3+n)}$
&\qquad& $(\alpha t)^{2/(3+n)}$ &\qquad& $(\sigma_{\psi}t)^{1/2}$ \\ \hline 
$E(t)$ && $(\sigma_{\psi}t)^{-1}$ && $\alpha^2 \Ki^{n+1}$ &&
$(\alpha t)^{2(n+1)/(3+n)}$ && $(\sigma_{\psi}t)^{-1}$ \\ \hline 
\end{tabular}
\caption{Division of domains of spectral exponent~$n$ according to 
predicted typical scale and energy of solutions
to Burgers equation as depending on time $t$.}
\label{t:n-Burgers-division}
\end{table}
Here we take into account that the increments of the potential in (\ref{MAX}) is
$(\psi_0(x)-\psi_0(0))\sim \alpha x^{(1-n)/2}$ for $n<1$ and is $\sim
\sigma_{\psi}$ for $n>1$ \cite{Gurbatov,Vergassola}. We assume further that
$\Ki=0$ for $n<1$ and that there is some cutoff number $\Ku$ for $n>1$. In the
range $n<-3$ we also need to have $\Ki>0$ as the solution of Burgers equation
exists only if the potential grows slower than quadratically (see the maximum
representation~(\ref{MAX})) and this implies that the spectrum must be
shallower than~$k^{-3}$ when~$k\to 0$.  From the equation~(\ref{velocity111}),
we have that at large time between the shocks $x_{{\rm shock},m}<x<x_{{\rm
shock},m+1}$ the velocity field has an universal structure $v (x,t) =
(x-a_m)/t$ and so the energy of Burgers turbulence my be estimated as $E(t)
\sim L^2(t) / t^2$ (see Table~\ref{t:n-Burgers-division}).  For the energy to
be finite in the range $-3<n<-1$, we require that there is some cutoff
wavenumber $\Ki>0$.  It has been known for some time that the behaviour of
$L(t)$ and $E(t)$ in $1<n$  has logarithmic corrections
\cite{Kida,GS81,FF,GSAFT97}.

If indeed Burgers turbulence is characterized by a single scale
$L(t)$, by dimensional analysis the spectrum takes the following 
self-similar form:
\begin{equation}
E(k,t) = \frac{L^3(t)}{t^2} \tilde E(kL(t)),
\label{ssspectr}
\end{equation}
It is well known that for an initial spectrum with $n>2$ 
the parametric pumping of energy to the area at small $k$'s
leads to the universal quadratic law, $E(k,t) \sim k^2$,
and  for $n<2$ we have conservation of
initial spectrum  $E(k,t)=E_0(k)= \alpha^2 |k|^n $ 
at small wavenumber, which is the spectral form of principle of ``permanence of large
eddies'' (PLE) \cite{Frisch,GSAFT97}.
In Fourier space the self-similarity ansatz (\ref{ssspectr}),
together with the PLE gives the same relations for integral scale
and the energy as written above, but now in the 
region $-3<n<2$.
Clearly, this argument cannot be
applied with initial data such that the spectral index $n\ge 2$, since
the later spectrum has now a $k^2$ dependence at small $k$ with a {\em
time-dependent\/} coefficient. But comparing this with Table 2, where 
the validity of $L(t) \sim (\alpha t)^{2/(3+n)}$
is $n<1$, we see 
that the region $1<n<2$ has to be a case apart.
In the interval $1<n<2$ the self-similarity ansatz is not
correct, as was shown in \cite{GSAFT97}. The reason for
this is competition between the initial~$|k|^n$ (with constant
prefactor~$\alpha^2$), and the autonomously generated 
$k^2$ (with prefactor increasing in time). If $n>2$, the initial spectrum,
at low $k$, is soon overwhelmed by $k^2$ generated by nonlinear interactions
between harmonics. In this case, hence, the spectrum is fully
universal, characterized by a single scale $L(t)$, and
otherwise independent of spectral index $n$. 

For sufficiently large wave numbers, the spectrum should always be dominated
by shocks. In one range, we should therefore have
\begin{equation}
E(k,t) \sim  \frac{B(t)}{k^2} \qquad\qquad\hbox{$k$ large}
\label{Elargek}
\end{equation}
which is equivalent to (\ref{ssspectr}) if $B(t) \sim L(t)/t^2$.
The amplitude of the small scale part of the spectrum will decrease with time
for $n>-2$ and increase with time for $-3<n<-2$.

We note that the spectrum gives only partial information on the
solutions to Burgers equation. Indeed, a $k^{-2}$ tail does not
distinguish between discontinuous solutions with shocks, and 
standard Brownian motion, which is almost surely continuous.
See \cite{SheAurellFrisch,Sinai,Vergassola} for other
characteristics of the mass and velocity distribution.

The rest of the paper will establish the regions in $k$ and $t$ for which the
above tables is true if we have cutoff at large and small wave-numbers.

\section{Homogeneous velocity and homogeneous potential;
$(n>1)$ and $(n<-3)$}.
\label{s:phivel} 

In this chapter we consider the evolution of Burgers turbulence in the region
$n>1$ assuming that both velocity and  potential are homogeneous Gaussian 
random function.  It means that we necessarily have some ultraviolet cutoff
wavenumber $\Ku$. The function  $b_0(k)$ can be characterized by a wavenumber
$\Ku$ around which lies most of the initial energy and which is, in order of
magnitude, the inverse of the initial integral scale $L_0$.  Similar situation
we have for the spectrum $n< -3$ and cutoff~$\Ki$ at small wave number.

Most of the  results  about the energy decay for this region have already been
obtained by Kida \cite{Kida}, but for the discrete model of initial condition. 
He introduced a model of discrete independent potential values in adjacent
cells, while their relation to the properties of the initial conditions (say,
the spectrum) were left unspecified. For the case of a p.d.f. with a Gaussian
tail, he obtained the functional form for the correlation function, energy
spectrum and the log-corrected $1/t$ law for the energy decay  $E(t)\sim t^{-1}
\sigma_{\psi}\ln^{-1/2}\left(t/\tnl\right)$, where however in the definition of
the nonlinear time $\tnl$ was some free parameter - the length of cell in the
discrete model.  In the more recent contributions~\cite{GS81},\cite{FF} (see
also \cite{Gurbatov}) the authors conjectured the asymptotic existence of a
Poisson process. This was then proved in \cite{Molchanov}, showing that, in the
$x$-$\psi$ plane, the density of the points is uniform in the $x$-direction and
exponential in the $\psi$-direction. This permits the calculation of the one-
and two-point p.d.f.'s of the velocity \cite{GS81} and also the full $N$-point
multiple time distributions (\cite{Molchanov}.  In papers \cite{GS81},\cite{FF}
was also shown that the statistical properties of the points of contact between
the parabola and the initial potential can be obtained from the statistical
properties of their intersections, whose mean number can be calculated using
the formula of Rice \cite{LeadbetterLindgrenRootzen}. Thus, it is possible to
express the parameters in the asymptotic formulas in terms of the r.m.s.
initial potential and velocity.

In the limit of vanishing viscosity, as the time $t$
tends to infinity, the statistical solution becomes self-similar 
and the energy spectrum has the form (\ref{ssspectr}). 
The integral scale $L(t)$
and the energy $E(t)$ are given, to leading order, by
\begin{eqnarray}
\label{Ltasymptotic}
L(t)&\simeq& (t\sigma_{\psi})^{1/2}\ln^{-1/4}\left({t\over
2\pi\tnl}\right)\ ,\\
E(t)&\simeq& t^{-1} \sigma_{\psi}
\ln^{-1/2}\left({t\over 2\pi\tnl}\right)\ ,
\label{Etasymptotic}
\end{eqnarray}
where
\begin{equation}
\tnl\equiv L_0^2/ \sigma_{\psi} =  L_0/ \sigma_v,
\qquad L_0\equiv \sigma_{\psi} /\sigma_v,
\label{deftnL0}
\end{equation}
are the nonlinear time and  the initial integral scale of turbulence.
Using this definition we can rewrite in a  first approximation 
\begin{equation}
L(t)\simeq L_0(t/\tnl)^{1/2},\qquad E(t)\simeq E_0 (t/\tnl)^{-1}.
\label{Ltf}\\
\end{equation}
The non-dimensionalized self-similar correlation function $\tilde
B_v(\tilde x)$, which is a function of
$\tilde x =x/L(t)$, is given by
\begin{equation}
\tilde B_v(\tilde x)= {d\over d\tilde x}\left(\tilde xP(\tilde x)\right),
\label{Bvasymptotic}
\end{equation}
where for $x \geq 0$
\begin{equation}
P(\tilde x) = {1\over2}\int_{-\infty}^\infty {dz \over
g\left({\tilde x+z \over 2}\right) \exp\left[{(\tilde
x+z)^2\over8}\right] + g\left({\tilde x-z \over 2}\right) \exp\left[{(\tilde
x-z)^2\over8}\right]},
\label{defPnoshock}
\end{equation}
\begin{equation}
g(z)\equiv \int_{-\infty}^z e ^{-s^2/2}\,ds.
\label{deferror}
\end{equation}
Our choice of normalization of the energy as $E(t)=L^2(t)/t^2$ imposes
that for the dimensionless spectrum 
we have $\int\tilde E(\tilde k)\, d \tilde k = 1$. It may be
shown that the function $P(\tilde x)$ is the probability of  having no
shock within an Eulerian interval of length $\tilde x L(t)$
\cite{Gurbatov}.

Note that the properties of the self-similar state are universal in so
far as they are expressed solely in terms of two integral
characteristics of the initial spectrum, namely the initial
r.m.s.~potential $\sigma_{\psi}$ and r.m.s.~velocity $\sigma_v$. Observe
that the spectral exponent does not directly enter, in contrast to
what happens when $n<1$ (see sections \ref{s:psi},\ref{s:nonpsi}).
For the dimensionless spectrum 
\begin{eqnarray}
\tilde E(\tilde k) &=& \frac{1}{2\pi} \int_{-\infty}^{\infty}
\tilde B_v(\tilde x) \exp(i\tilde k \tilde x)\, d \tilde x \nonumber \\
&=& -\frac{ik}{2\pi} \int_{-\infty}^{\infty}
\tilde x P(\tilde x) \exp(i\tilde k \tilde x)\, d \tilde x,
\label{dimlesspec}
\end{eqnarray}
we have the following asymptotic:
\begin{equation}
\tilde E(\tilde k)
\simeq \left\{ 
\begin{array}{rl}
a_{+} \,{\tilde k}^2,    & \tilde k \ll 1 \\[1.5ex]
a_{-} \,{\tilde k}^{-2}, & \tilde k \gg 1 \\
\end{array}
\right., \quad \tilde k \equiv kL(t)\ .
\label{tEasympt}
\end{equation}
The $k^{-2}$ region is the signature of shocks, while the
$k^2$ region comes due to the parametric pumping of 
energy to the area of small $k$.  The two constants~$a_{+}$ and~$a_{-}$
can be computed theoretically as~\cite{Gurbatov}
\begin{eqnarray}
a_{+} &=& 1/\pi \int_0^\infty \tilde x^2 P(\tilde x)\, d\tilde x = 1.078\ldots
 \nonumber\\
a_{-} &=& 2 \pi^{-3/2} = 0.359\ldots
\end{eqnarray}

In dimensioned variables, the small-$k$ region behavior of the spectrum is thus 
\begin{equation}
E(k,t) = a_{+}\frac{L^5(t)}{t^2} \,k^2 = A(t) \,k^2,\quad kL(t)\ll
1,
\label{Eleft}
\end{equation}
where
\begin{equation}
A(t) \simeq a_{+}\sigma_v^2 L_0^3 \left(\frac{t}{\tnl } \right)^{\frac{1}{2}}
\,
\ln^{-5/4} \left(\frac{t}{2 \pi \tnl }\right).
\label{Aoft}
\end{equation}
So, we have a spectrum with an algebraic $k^2$ region and a time-increasing
coefficient $A(t)$.

The situation is more complicated at large but finite time \cite{GSAFT97}.  We
must now  distinguish two cases.  When $n>2$, the $k^2$ contribution
(\ref{Eleft}) dominates everywhere over the $|k|^n$ contribution and we have a
self-similar evolution in the whole range of wavenumbers (see
Fig.~\ref{fig:summary}~a), but the ``self-similar'' time $t_{\rm ss}$ from
which we have self-similar  stage of evolution depends on $n$. In the general
case, the condition $t/\tnl \gg 1$ is not enough for the  Poisson approximation
to hold and consequently for the existence of self-similarity.  Let us denote
by $\Delta_{\rm corr}$ a typical correlation length for the initial potential,
which may be greater the initial integral scale $L_0$.  The self-similarity
occurs when the integral scale of the turbulence $L(t)$ (\ref{Ltasymptotic}) is
much greater the typical correlation length $\Delta_{\rm corr}$. This leads to
the following condition on the ``self-similar'' time $t_{\rm ss}$ \cite{GSAFT97}
\begin{equation}
t_{\rm ss} \sim \tnl 
\left(\frac{\Delta_{\rm corr}}{L_0}\right)^2 \ .
\label{c1dominanslocal}
\end{equation}
There are instances where $(\Delta_{\rm corr}/L_0)^2$ can be large.  Consider
an initial spectrum $E_0(k)$ (\ref{initsp}) with $n \gg 1$ and a function 
$b_0(k)$ decreasing rather fast when $k>\Ku$.  In this case the initial
velocity field is a quasi-monochromatic signal with a center wavenumber
$\Ku\sim L_0^{-1}$ and a width $\Delta k\sim \left[\Delta_{\rm
corr}\right]^{-1} \ll \Ku$. At the early stage of evolution $\tnl \ll t \ll
\tnl(\Delta_{\rm corr}/L_0)$ we have the saturation of amplitude modulation and
the shift of the shocks is much smaller then the period of the
quasi-monochromatic signal.  The energy of this signal is approximately the
same as the energy of the periodic wave\,: $E(t) \simeq L_0^2/12t^2$
\cite{GurbatovMalakhov1977}. Nevertheless due to the finite width of the
initial spectrum, we have the generation of a low frequency component
$v_l(x,t)$ whose spectrum is well separated from the primary harmonic $\Ku$ and
with the energy $E_l(t) \sim E_0(L_0/\Delta_{\rm corr})^2 \ll E_0$.  At
$\tnl(\Delta_{\rm corr}/L_0) \ll t \ll \tnl(\Delta_{\rm corr}/L_0)^2$ the
energy of the low frequency component is larger than the energy of the
high-frequency quasi-periodic wave, but to the large spatial scale we have a
relatively small distortion of this component. And only at  $t \gg t_{\rm
ss}\sim \tnl (\Delta_{\rm corr}/L_0)^2$ do we have the self-similar regime of
evolution.  The physical reason for this is a strong correlation of the shocks
in the early stage of the evolution, which prevents the rapid merging of
shocks. We need to stress that we have a similar situation for a spectrum $n
\ll -3$ and a cutoff~$\Ki$ at small wave number.

\begin{figure}[tb]
 \centerline{\epsfig{file=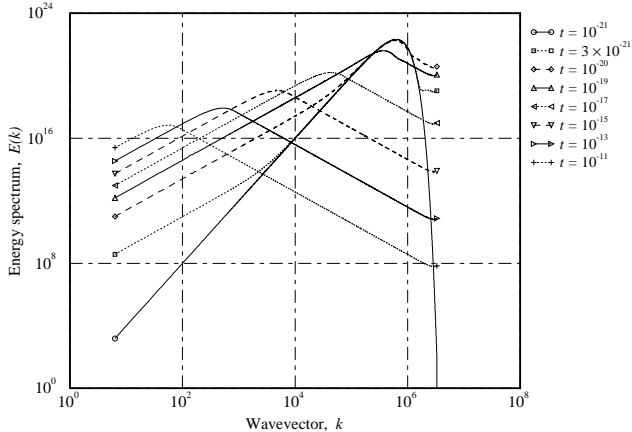,width=6cm}}
 \vspace{6mm}
 \caption{Evolution of the energy spectrum with an initial spectrum
proportional to $k^4$ at small wavenumbers $k$. The spatial resolution is $N=2^{20}$.
Spectra averaged over 3000~realizations. The labels correspond to
output times $t/\tnl=0.033$,\ldots, up to~$t/\tnl=3.3 \times 10^8$.}
 \label{figb-4n}
\end{figure}

\begin{figure}[tb]
 \centerline{\epsfig{file=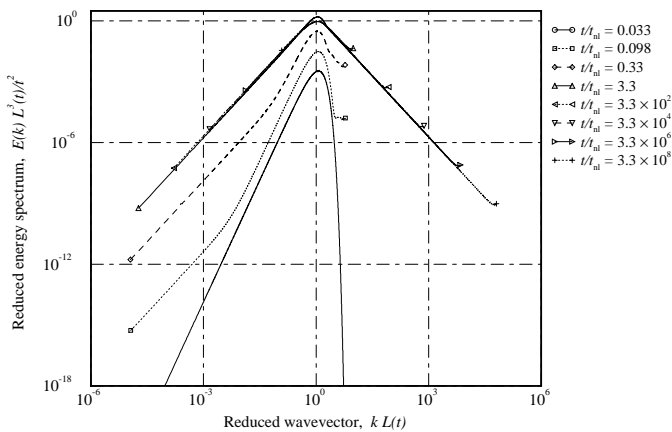,width=6cm}}
 \vspace{6mm}
 \caption{Reduced energy  
spectra $E(k,t)t^2/L_{\rm exp}^3(t)$ ($n=4$) at the same times as a function of 
reduced wave number $k L_{\rm exp}(t)$.}
 \label{figb-4s}
\end{figure}

Let us now discuss the results of numerical simulations.  We use a smooth
cutoff of the initial power spectrum~(\ref{NIS}) with $\Ku=N/16$. We consider
in all experiments periodic initial condition, so the infra-red cutoff
frequency in this case is determined by the size of simulation box and
$\Ki=2\pi$. To check the self-similar ansatz we consider the evolution of
energy spectrum $E(k,t)$, of the energy $E(t)=\langle v^2(x,t)\rangle$ and of
the integral scale $L_{\rm exp}(t)$ which we can measure from the experimental
data as
\begin{equation}
L^2_{\rm exp}(t)=\langle \psi^2(x,t)\rangle/\langle v^2(x,t)\rangle \ .
\label{Lexp}
\end{equation}

In Fig.~\ref{figb-4n} energy spectra (averaged over about $3000$~realizations
of the random process) are shown at different moments of time from
$t/\tnl=0.033$ to $t/\tnl=3.3 \times 10^8$.  The initial spectrum was $k^4$ at
small $k$.  Fig.~\ref{figb-4s} contains  reduced energy spectra
$E(k,t)t^2/L_{\rm exp}^3(t)$ at the same time as a function of reduced wave
number $k L_{\rm exp}(t)$.  We see the generation of $A(t)\,k^2$ with growing
amplitude $A(t)$ at small $k$, and $k^{-2}$ at large wavenumber. The switch
point $\Ks(t)$ between $A(t)k^2$ and $\alpha^2 |k|^n$ regions of the spectrum
moves quickly towards the maximum of the spectrum and finally the $\alpha^2
k^4$ part of the spectrum disappears (see Fig.~\ref{fig:summary}a).


The preservation of the shape of each curve at large time $t/\tnl > 10$  
is evident. It is also seen
that the total energy (the area under the curves) decreases, and so does
the characteristic wavenumber $k_L(t) \sim 1/L(t)$.  The curves in
Fig.~\ref{figb-4s} have been plotted with respect to the reduced
wavenumber~$k\,L_{\rm exp}(t)$ as the quantity~$L_{\rm exp}(t)$ can be measured
unambiguously from the numerical simulations.  To compare the results
with the spectrum theoretical $\tilde E(\tilde k)$ (\ref{dimlesspec}),
we need to deduce the relation between~$L_{\rm exp}(t)$ and~$L(t)$, that we can
get from~(\ref{ssspectr})
\begin{eqnarray}
L^2_{\rm exp}(t) &=& L^2(t)
\frac{\int_{-\infty}^{\infty} \tilde E(\tilde k)\tilde k^{-2}\,d \tilde k}
{\int_{-\infty}^{\infty} \tilde E(\tilde k)\,d \tilde k} \nonumber \\
&=& L^2(t) \int_0^\infty \tilde xP(\tilde x)\, d\tilde x \approx 1.65\ L^2(t).
\label{LexpL}
\end{eqnarray}

From these, we deduce the experimental values of the constants~$a_{+}=1.10$
and~$a_{-}=0.37$, both slightly larger ($\approx$\,2\,\%) than the theoretical
values.  This very small discrepancy could be due to finite-size effects
contaminating the measurement of~$\langle \psi^2(x,t)\rangle$ at small
wavenumbers and~$\langle v^2(x,t)\rangle$ at large wavenumbers, and thus of the
experimental integral scale~$L_{\rm exp}(t)$.



The asymptotic spectrum is reached rather quickly after the non-linear
time~$\tnl$ and the asymptotic formula describes the numerical data very well,
not only in the limits of relatively large and small wave numbers, but also at
the top, where the spectrum switches between the two asymptotes.  From
Fig.~\ref{figb-4s} is also seen that the transition between the two asymptotics
$\tilde k^2$ and $\tilde k^{-2}$ is rather sharp.



\subsection{Breakdown of self-similarity; $(1<n<2)$}
\label{s:break}

The  more interesting case is $1<n<2$, when we have breakdown of the 
self-similarity \cite{GSAFT97}.
The permanence of large eddies implies now that, at extremely small $k$,
\begin{equation}
E(k,t) \simeq \alpha^2 |k|^n,\quad  {\rm for}\,\, k\to 0.
\label{Eksmall}
\end{equation}
This relation holds only in an outer region $|k|\ll \Ks(t)$ where
(\ref{Eksmall}) dominates over (\ref{Eleft}). The switching wavenumber
$\Ks(t)$, obtained by equating (\ref{Eleft}) and (\ref{Eksmall}), is
given by
\begin{eqnarray}
\Ks(t) &\simeq& \left(\frac{\alpha^2 t^2}{L^5(t)}\right)^{\frac{1}{2-n}}
\nonumber \\
&\simeq& L_0^{-1}
\left(\frac{t}{\tnl}\right)^{-\frac{1}{2(2-n)}}
\ln^{\frac{5}{4(2-n)}} \left(\frac{t}{2 \pi \tnl}\right).
\label{ksasympt}
\end{eqnarray}
Let us define an energy wavenumber $k_L(t) =
L^{-1}(t)$, which is roughly the wavenumber around which most of the
kinetic energy resides. From (\ref{Ltasymptotic}) $k_L(t) \sim
(t\sigma_\psi)^{-1/2}$ (ignoring logarithmic corrections). We then
have from (\ref{ksasympt}), still ignoring logarithmic corrections\,:
\begin{equation}
\frac{\Ks(t)}{k_L(t)} \sim
\left(\frac{t}{\tnl } \right)^{- \frac{n-1}{2(2-n)}}.
\label{kskLratio}
\end{equation}
Hence, the switching wavenumber goes to zero much faster than the
energy wavenumber, so that the preserved part of the initial spectrum~$|k|^n$
becomes rapidly irrelevant.
Let us also observe that the ratio of the energy in the outer region
to the total energy, a measure of how well the Kida law (\ref{Ltasymptotic})
is satisfied,
is equal to $(t/\tnl)^{-3(n-1))/(n-2)}$ (up to logarithms) and thus
becomes very small when $t\gg \tnl$, unless $n$ is very close to unity.
Thus, for $1<n<2$ there is no globally self-similar evolution of the energy
spectrum at finite time. Of course, as $n\to 2$ the inner $k^2$ region overwhelms the
outer $|k|^n$ region and as $n\to 1$ the converse happens, so that in both
instances global self-similarity tends to be reestablished.

\begin{figure}[tb]
 \centerline{\epsfig{file=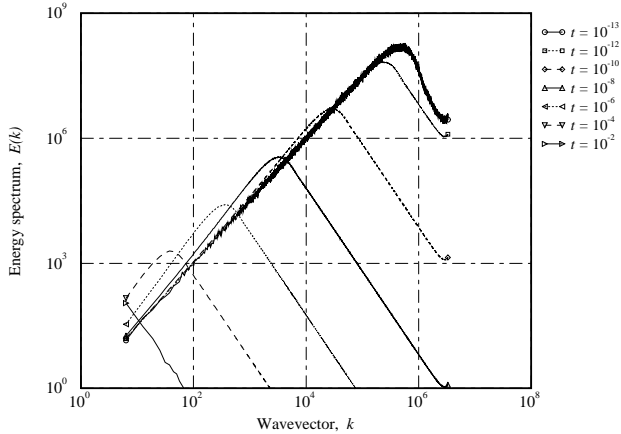,width=6cm}}
 \vspace{6mm}
 \caption{Evolution of the energy spectrum with an initial spectrum
proportional to $|k|^n$ ($n=1.5$) at small wavenumbers
$k$. Resolution $N=2^{20}$.  The labels correspond to output times
$t_1/\tnl=0.18$,\ldots, up to~$t/\tnl = 1.8 \times 10^{10}$.}
 \label{figb-1.5n}
\end{figure}


\begin{figure}[tb]
 \centerline{\epsfig{file=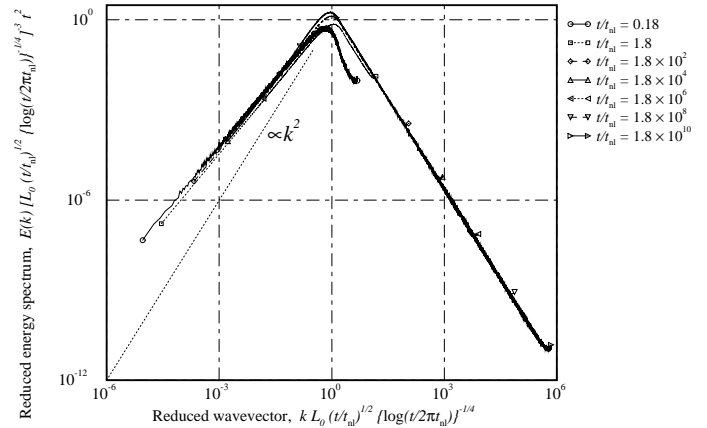,width=6cm}}
 \vspace{6mm}
 \caption{Reduced energy  
spectra $E(k,t)t^2/L^3(t)$ at the same times as a function of 
reduced wave number $k L(t)$ using the theoretical value
$L(t)= (t/\tnl)^{1/2}[\log(t/2\pi \tnl)]^{-1/4}$. }
 \label{figb-1.5l}
\end{figure}

In Fig.~\ref {figb-1.5n} energy spectra (averaged over $\sim 3000$~realizations
of the random process) are shown at different moments of time from
$t/\tnl=0.18$ to $t/\tnl=1.8 \times 10^{10}$.  For the  initial spectrum we
have  $n=1.5$.  Fig. ~\ref{figb-1.5l} contains  reduced energy spectra
$E(k,t)t^2/L^3(t)$ at the same time as a function of reduced wave number $k
L(t)$, when  we use the asymptotic expression~(\ref{Ltasymptotic}) for $L(t)$. 
We see again the generation of $A(t)\,k^2$ with growing amplitude $A(t)$ at
small $k$ and $k^{-2}$ at large wavenumber.  The switch point $\Ks(t)$ between
$A(t)\,k^2$ and $\alpha^2 |k|^n$ regions of the spectrum tends slowly to the
origin of the spectrum and finally the $\alpha^2 |k|^n$ part of the spectrum
disappears (see Fig.~\ref{fig:summary}~b).

\begin{figure}[tb]
 \centerline{\epsfig{file=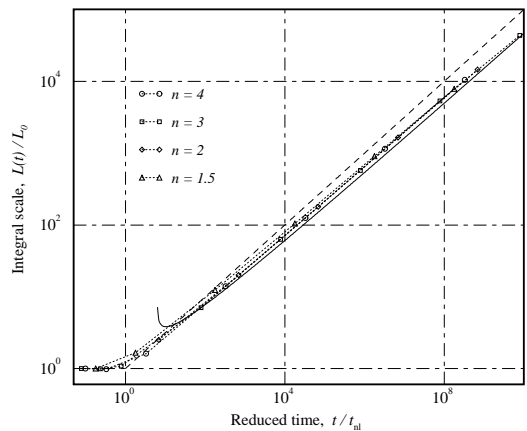,width=6cm}}
 \vspace{6mm}
 \caption{Evolution
of the computed integral scale $L_{\rm exp}(t)$ for $n=1.5, 2,3,4$
(dotted lines) compared to the
theoretical leading-order prediction $\tilde {L}_{\rm exp}=1.28\,L(t)$
(\ref{Ltasymptotic}),(\ref{LexpL}) (solid line) and the same without
the logarithmic correction (dashed line).}
 \label{figlb-2pp}
\end{figure}

\begin{figure}[tb]
 \centerline{\epsfig{file=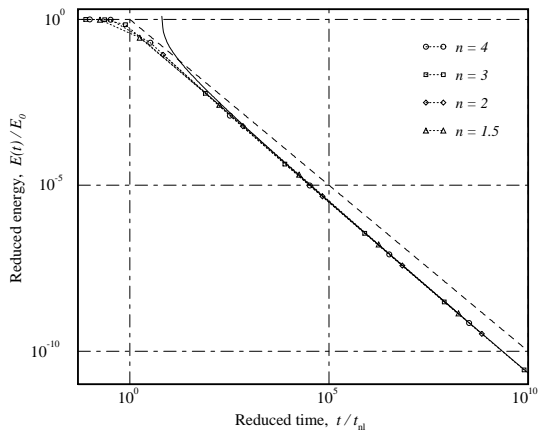,width=6cm}}
 \vspace{6mm}
 \caption{Evolution
of the computed integral energy  $E(t)$ for $n=1.5, 2,3,4$ (dotted lines)
compared to the
theoretical leading-order prediction $E(t)=L^2(t)/t^2$ (\ref{Etasymptotic}) 
(solid line) and the same without
the logarithmic correction (dashed line).}
 \label{figeb-2p}
\end{figure}

Thus the numerical experiments support the theoretical prediction that for
$n>1$ at large time we have the self-similar behavior of the spectrum
(\ref{ssspectr}). Moreover the integral scale of turbulence
$L(t)$(\ref{Ltasymptotic}) and the energy of turbulence $E(t)$  are in perfect
agreement with theoretical predictions, even for $1<n<2$ where we have breaking
of self-similarity.  On Fig. \ref{figlb-2pp} and \ref{figeb-2p}  we plot the
evolution of $L_{\rm exp}(t)$ and $E(t)$ for different $n=1.5, 2, 3, 4 $ and
the theoretical curves, taking into account the relation between $L_{\rm
exp}(t)$ and $L(t)$ (\ref {LexpL}), showing that the theoretical predictions
are perfectly reproduced by the simulations.

In all experiments, we only consider times small enough for the characteristic
wavenumber $k_L(t) \gg \Ki$, so that we still have many shocks in the box. When
this condition is broken, it means we have a single shock in the
simulation domain and we have universal self-similar linear decay of the
spectrum $E(k,t) \sim (kt)^{-2}$.


\section{Homogeneous velocity  and non-homogeneous potential; $(-1<n<1)$}
\label{s:psi}

We begin with the case $-1<n<1$ when the initial potential has homogeneous
increments. Many aspects of this case are well understood, thanks in particular
to Burgers' own work \cite{Burgers}, who did consider the case when the initial
velocity is white noise $n=0$ (see also \cite{Gurbatov,WW98}).  

The phenomenology is quite simple (see section \ref{s:large-time decay}).
Increments $\Delta \psi_0(L((t))$ of the initial potential over a distance
$L(t)= x-\tilde{a}(x,t)$ can be estimated from the square root of the structure
function $S_{0\psi}(L)$ of the potential (\ref{defstr})
\begin{equation}
S_{0\psi}(x) \equiv\langle \left(\psi_0(x)-\psi_0(0)\right)^2\rangle,
\label{defstr}
\end{equation}
When  when $n<1$, it grows without 
bound $S_{0\psi}(x) \sim \alpha^2|x|^{1-n} $.  For a given
position $x$, the maximum in (\ref{MAX}) will come
from those $a$\ms1 such that the change in potential is
comparable to the change in the parabolic term and this immediately leads to
$L(t) \sim \alpha  t^{2/(n+3)}$. (see Table (\ref{t:n-Burgers-division}).

When the initial spectrum (\ref{initsp}) has no cutoff wavenumber, we can use
that scaling and then get immediately that the above mentioned expression for
the integral scale $L(t)$ and energy $E(t)$ (see
Table~\ref{t:n-Burgers-division}) are now exact for all time starting from
zero.  For details, see \cite{Vergassola} (Section~4) and \cite{AvellanedaE}. 
But even when the initial spectrum has both cutoff waves numbers  at large and
small scale there is some region in ($kt$) plane where we have self-similar
evolution of the spectrum.
 
In the general case for the structure function~(\ref{defstr}) of the potential,
we have
\begin{eqnarray}
 S_{0\psi}(x) &=& g(x)\beta^2 \alpha^2|x|^{1-n}\ , \nonumber \\
 \beta^2 &=& \alpha^2 {{2\pi}\over{\Gamma(2-n)\sin\frac{\pi(n-1)}{2}}} 
 \label{Scatoff}
\end{eqnarray}
The properties of the dimensionless function $g(x)$ are determined by the
function $b(k)$ and for $b(k)\equiv 1$ we have  $g(x)\equiv 1$. 
When the initial spectrum
has cutoff wavenumbers $\Ki$ and $\Ku$ the function $g(x)=1$ in some spatial
interval  $L_{\rm u} <x<L_{\rm i}$ where $L_{\rm u} \sim 1/\Ku$ and
$L_{\rm i} \sim 1/\Ki$.
Let us introduce the dimensionless variables
\begin{equation}
 \tilde{x}=x/L(t),
 \qquad \tilde{a}=a/L(t), \qquad \tilde{v}=v/(L(t)/t),
   \label{dimlessxav}
\end{equation}
where $L(t)$ is an integral scale of turbulence. 
Then the ``maximum representation'' will be rewritten in the form 
\begin{equation}
\tilde {v} (\tilde {x},t) = \tilde {x}-\tilde {a}(\tilde {x},t),
\label{velocss}
\end{equation}
where $\tilde{a}(\tilde{x},t)$ is the coordinate at 
which dimensionless function~$\tilde {\Phi}(\tilde {x},\tilde {a},t)$ achieves its
(global) maximum  for given $\tilde {x}$ and $t$ and
\begin{equation}
\tilde {\Phi}(\tilde {x},\tilde {a},t) = \tilde {\psi}_0(\tilde{a},t) 
- \frac{(\tilde {x}-\tilde {a})^2}{2}.
\label{Gfss}
\end{equation}
here $\tilde {\psi}_0(\tilde{a},t)=\psi_0(\tilde{a}L(t))\left(t/L^2(t)\right) $ is 
the dimensionless initial potential with
the following structure function
\begin{equation}
  \tilde {S}_{0\psi}(\tilde{x})=g(\tilde{x}L(t)) \beta^2 |\tilde{x}|^{1-n} 
   \label{Strss}
\end{equation}
Here we define that integral scale~$L(t)$ by the relation
\begin{equation}
  L(t)= (\alpha t)^{2/(3+n)} 
  \label{Scatoffss}
\end{equation}
and so in (\ref{Strss}) we use that $(\alpha t)^2/L^{(3+n)}(t)=1$.
For this definition of the integral scale, we have for the dimensionless
spectrum (\ref{ssspectr})
\begin{equation}
\tilde E(\tilde k)
= \left\{ 
\begin{array}{rl}
{\tilde k}^n,    & \tilde k \ll 1 \\[1.5ex]
\gamma_n{\tilde k}^{-2}, & \tilde k \gg 1 \\
\end{array}
\right., \quad \tilde k \equiv kL(t).
\label{Ekofn}
\end{equation}
and for the energy of Burgers turbulence 
\begin{equation}
  E(t)=a_n L^2(t) / t^2=a_n \alpha^{\frac{4}{3+n}} t^{-\frac{2(n+1)}{3+n}} 
\label{Etofn}
\end{equation}
where $\gamma_n$ and $a_n$ are dimensionless constants, which we will
determine from the numerical experiments.

Consider first the case when we have a cutoff wavenumber~$\Ku$ only at small
scale ($\Ki=0$). For $L(t)\gg L_{\rm u}$ the structure function
(\ref{Scatoffss}) my be replaced by
\begin{equation}
  \tilde {S}_{0\psi}(\tilde{x})= \beta^2 |\tilde{x}|^{1-n}, 
   \label{Sss}
\end{equation}
the function not depending on time and possessing no spatial scales in its own.
In this case the statistical properties of absolute maxima coordinates $\tilde {a}$
do not vary with time. The latter phenomenon means that the 
field $\tilde {v}(\tilde {x},t)$ (\ref{velocss}) statistical properties 
determined by time-independent statistics of  $\tilde {a}(\tilde {x},t)$ 
are rendered self-similar according (\ref{velocss}).
So at large times when  $L(t)$ is also large, we have self-similar evolution
of the spectrum Eq.(\ref{ssspectr}) (see Fig.~\ref{fig:summary}~c).
Alternatively, we could argue that when $t$ is so large that the
parabolas appearing in (\ref{MAX}) have a radius of curvature
much larger than the typical radius of curvature of features in the
initial potential, we can plausibly replace that initial potential by
fractional Brownian motion of exponent $h=(1-n)/2$, so that
the upper cutoff~$\Ku$ becomes irrelevant. Without loss of generality we
may assume that this fractional Brownian motion starts at the origin
for $x=0$.  This function is then statistically invariant under the
transformation $x \to \lambda x$ and $\psi_0 \to \lambda ^h\psi_0$. It
is then elementary, using (\ref{MAX}), to prove
that a rescaling of the time is (statistically) equivalent to a
suitable rescaling of $x$ distances and of $\psi(x,t)$. This implies
 the above expression for $L(t)$ and $E(t)$ (see Table \ref{t:n-Burgers-division}).

For the self-similar initial spectrum with $-1<n<1$ we have a divergence
of the potential in the infrared part of the spectrum and divergence of
velocity and gradient at small scale. So  we can not introduce now the
initial scale on the base of (\ref{deftnL0}). Assuming that we have  
a cutoff wavenumber at small scale, we can in this section use
some other definition of nonlinear time and initial scale 
\begin{equation}
\Tnl =  L^{'}_0/ \sigma_v=1/\sigma_q,
\qquad L^{'}_0\equiv \sigma_{v} /\sigma_q,
\label{deftnLnew}
\end{equation}
It is easy to see that this time is equal to the the characteristic time of
shock formation $t_{\rm s}$.
Using this definition we can rewrite the expression for  $L(t)$ and $E(t)$  
in the form
\begin{eqnarray}
L(t)&\simeq& L^{'}_0(t/\Tnl)^{2/(3+n)},
\label{Ltfnew}\\
E(t)&\simeq& E_0 (t/\Tnl)^{-2(n+1)/(3+n)}.
\label{Etfnew}
\end{eqnarray}

Assume now that we have also a cutoff wavenumber $\Ki$ at large scale and we
have saturation of the potential structure function to $2\sigma^2_{\psi}$ at $x\gg L_{\rm
i}$.  In this case in the interval when $L_{\rm i}\gg L(t)\gg L_{\rm u}$ we 
can replace the dimensionless structure function (\ref{Scatoffss}) by
(\ref{Sss}). It means that in some time interval $t_{\rm u}\ll t  \ll t_{\rm
i}$, where $t_{\rm u}$, $t_{\rm i}$ are determined by the condition $L(t_{\rm
i})=L_{\rm i}$,  $L(t_{\rm u})=L_{\rm u}$, we will still have self-similar laws
for the integral scale $L(t)$ and energy $E(t)$ of turbulence.  The energy
spectrum $E(k,t)$ will have the self-similar behavior on this time in the
region $k>\Ki$ (see Fig.~\ref{fig:summary}~d). For the final state at very
large times, there are two possible situations.  In the first one, if we have a
finite box of size $L_{\rm box}=L_{\rm i}=2\pi/\Ki$ and a periodic initial
perturbation  with this period, then at very large time we will finally have
one triangular wave on the period and the energy will be decay as
$E(t)=L^2_{\rm i}/12t^2$. In the non-periodic case with a continuous power
spectrum, we will have the generation of a low frequency component $E(k,t) \sim
A(t)\,k^2$ in region $k<\Ki$ and finally the behavior of the turbulence will be
like in the case $n>1$ (see section \ref{s:phivel}).

In numerical simulations we used two models of initial spectrum. In the first
one we assume that we have the self-similar  power spectrum (\ref{initsp}) in
the whole wavenumber range from $2\pi$ to $N\pi$ ($\Ki=2\pi$, $\Ku=N\pi$), in
the second we use an infrared cutoff wavenumber $2\pi \ll \Ki \ll \Ku=N\pi$. 
If the initial spectrum is self-similar inside the interval
$\Ki=2\pi$,$\Ku=N\pi$ then for the nonlinear time $\Tnl$ we have
\begin{eqnarray}
 \Tnl &=& \left(\frac{n+3}{2}\right)^{1/2}
 \frac{1}{\alpha (\Ku^{(3+n)/2}-\Ki^{(3+n)/2})} \nonumber \\
 &\simeq& \left(\frac{n+3}{2}\right)^{1/2}\frac{1}{\alpha \Ku^{(3+n)/2}} \ .
 \label{deftnLss}
\end{eqnarray}

\begin{figure}[tb]
 \centerline{\epsfig{file=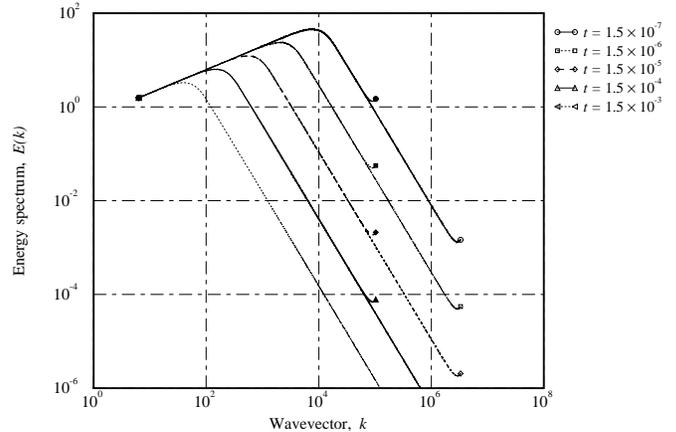,width=6cm}}
 \vspace{6mm}
 \caption{Evolution of the energy spectrum with an initial spectrum
proportional to $|k|^n$ ($n=0.5$) at small wavenumbers
$k$.  Resolutions $N=2^{15}$ and $N'=2^{20}$ are denoted by filled or open
symbols respectively. 
The labels correspond to output times $t_1=1.5 \cdot 10^{-7}$,\ldots, up to
$t=1.5 \cdot 10^{-3}$ (in absolute time, the reduced time~$t/\Tnl$ depends on
the value of the upper cutoff~$\Ku$, equal in this case to~$N\pi$ or $N'\pi$).
}
 \label{figh-.75n}
\end{figure}

\begin{figure}[tb]
 \centerline{\epsfig{file=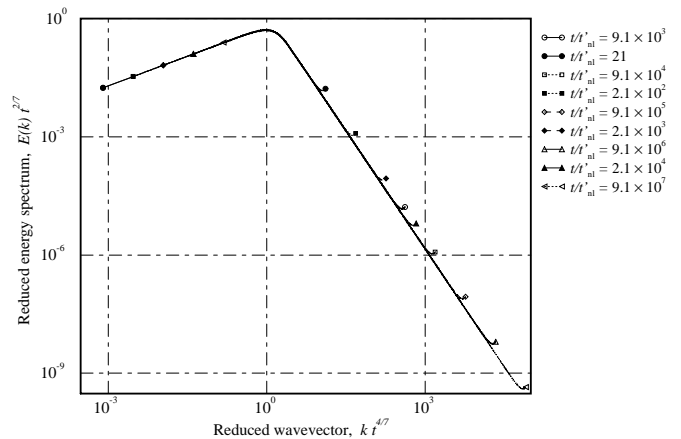,width=6cm}}
 \vspace{6mm}
 \caption{Reduced energy  
spectra $E(k,t)\,t^{2/7}$ ($n=0.5$) at the same times as a function of 
reduced wave number $k \,t^{4/7}$.  Same conditions as
in~Fig.~\ref{figh-.75n}.}
 \label{figh-.75t}
\end{figure}

In Fig.~\ref{figh-.75n} energy spectra $E(k,t)$ are shown at different moments
of time from $t/\Tnl = 21$ to $t/\Tnl=9.1\times 10^7$.  The initial spectrum was
$|k|^n$ with $n=0.5$, but two simulations were done with different
resolutions~$N=2^{15}$ and~$N'=2^{20}$, so that we have in reality two
different ultraviolet cutoffs~$\Ku=2^{15}\pi$ and~$\Ku'=2^{20}\pi$.  From the
figures, one can easily see that in the frequency range $\Ki<k<\min(\Ku,\Ku')$
the spectra are exactly equal to each other for all times. 
Fig.~\ref{figh-.75t} shows reduced energy spectra $E(k,t)\,t^{2/7}$ as a
function of reduced wave number $k \,t^{4/7}$ at different times for these two
different values of~$N$.  We see that once again, both spectra display a
perfectly self-similar evolution and are exactly equal in their common range of
reduced wavenumbers.  Moreover, it possible to show that we have not only the
conservation of the spectrum in presence of high frequency signal but also the
conservation of the large-scale structures in each unique realization
\cite{Aurell,GP99}.  The measured value of the dimensionless constant~$\gamma_{1/2}=1.62$ is reported in Table~\ref{t:n-constants}.

\begin{figure}[tb]
 \centerline{\epsfig{file=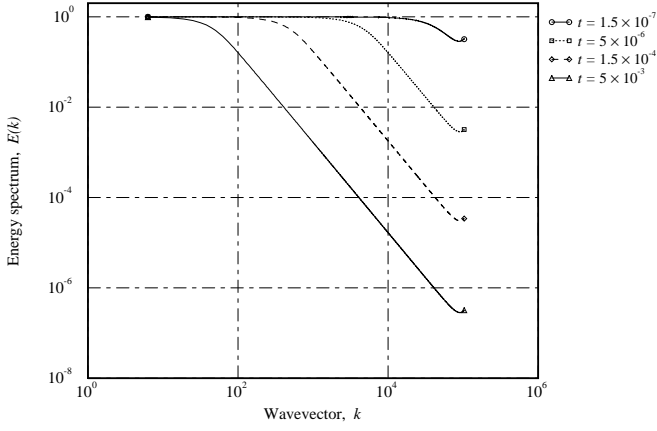,width=6cm}}
 \vspace{6mm}
 \caption{Evolution of the energy spectrum with an initial spectrum
independent of~$k$ ($n=0$) (white noise initial velocity).}
 \label{figh-.5n}
\end{figure}

\begin{figure}[tb]
 \centerline{\epsfig{file=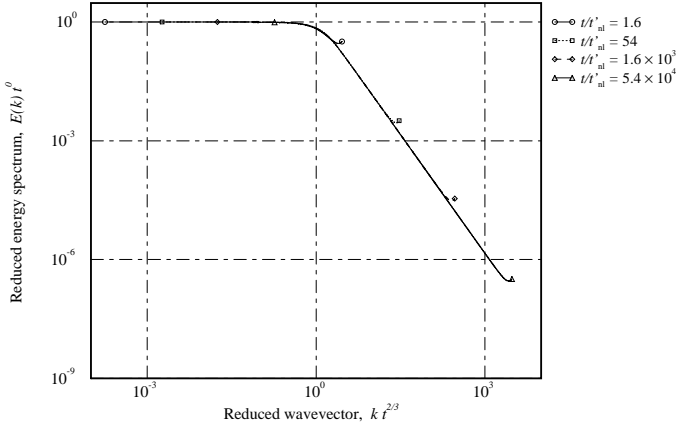,width=6cm}}
 \vspace{6mm}
 \caption{Reduced energy  
spectra $E(k,t)\,t^{0}$ ($n=0$) at the same times as a function of 
reduced wave number $k \,t^{2/3}$.}
 \label{figh-.5t}
\end{figure}

Similarly, in Fig.~\ref{figh-.5n}, energy spectra $E(k,t)$ are shown at
different moments of time from $t/\Tnl = 1.6$ to $t/\Tnl=5.4\times 10^4$ for
the initial spectrum was a classical white noise with $n=0$. 
Fig.~\ref{figh-.5t} contains  reduced energy spectra $E(k,t)\,t^0$  as a
function of reduced wave number $k \,t^{2/3}$ at five different time. The
observed value of the dimensionless constant~$\gamma_0=1.43$ is reported in
Table~\ref{t:n-constants}.



\begin{figure}[tb]
 \centerline{\epsfig{file=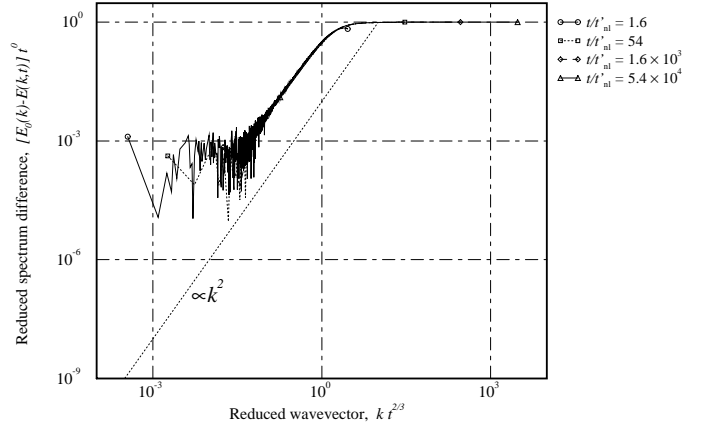,width=6cm}}
 \vspace{6mm}
 \caption{Evolution of the difference of the reduced energy spectrum with
the initial white noise spectrum~$k^0$, showing the universal~$k^2$ term
at small~$k$s, which is subleading in this case.}
 \label{figh-.5d}
\end{figure}

 From the structure  of the Burgers equation we see that due to nonlinear
interaction of harmonics we have always the $k^2$ term at low wave numbers $k$,
which may be leading or sub-leading.  The sign of this term depends on the
initial spectrum.  It is evident, that for $n>2$  we have generation of new
component at small wave number and this term increases with time see equation
(\ref{Aoft}).  When~$n<2$, this term is subleading, but its growing amplitude
can make it apparent so that it dominates the dynamics, as shown in
section~\ref{s:break}.  But when~$n<1$, this term is completely masked by the
initial components~$\sim |k|^n$, and the only way to show that it is really
present is by computing the diference of the spectrum with the initial one,
provided the statistical noise in the simulations is small enough.  This has
been done in Fig.~\ref{figh-.5d}), showing this subleading term $\sim k^2$,
this time with a negative amplitude, but displaying also perfect self-similar
behavior.

\begin{figure}[tb]
 \centerline{\epsfig{file=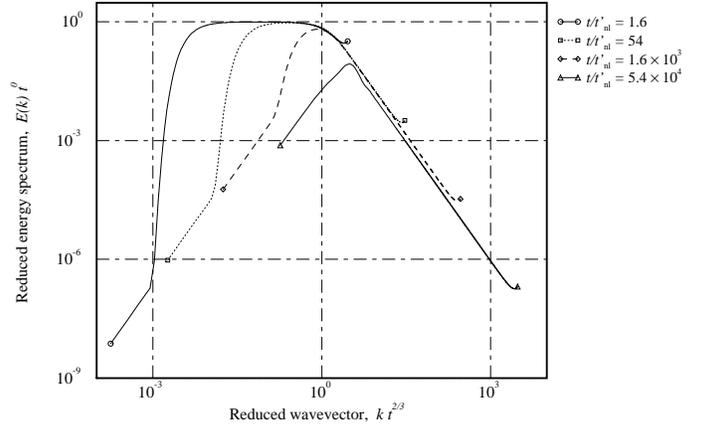,width=6cm}}
 \vspace{6mm}
 \caption{Evolution of the energy spectrum with an initial spectrum
independent of~$k$ ($n=0$) as a function of 
reduced wave number $k t^{2/3}$ with an initial infrared cutoff wavenumber at
$\Ki=64\,\pi$ (corresponding to $1/32$~of the box size).  One can see that the
upper part of the spectrum is unchanged with respect to Fig.~\ref{figh-.5t}.}
 \label{figsh-.5t}
\end{figure}

Another way to display this universal low wavenumber~$k^2$ component, is to
introduce some infrared cutoff in the initial condition.  In the simulation
shown in Fig.~\ref{figsh-.5t}, we consider the case of white noise initial
spectrum $n=0$ but with an infrared cutoff at $\Ki=64\,\pi$. For the first two
displayed times, we have a self-similar evolution of the spectrum in the
wave-number range $\Ki<k<\Ku$, but with the generation of a component spectrum
$E(k,t)\sim A(t)\,k^2$ in the region $k<\Ki$. At the time $t/\Tnl=1.6\cdot
10^3$, $L(t) \sim L_{\rm i}$ reaches the lower cutoff~$\Ki$, and the spectrum
then becomes equivalent and evolve in time as in section~\ref{s:phivel}, where
we have self-similar evolution with universal behavior of the spectrum 
$E(k)\sim k^2$ and $E(k)\sim k^{-2}$ at small and large wavenumber
respectively.  It is interesting to note that all parts of the spectrum~$k^2$,
$k^0$ and $k^{-2}$ evolve in a self-similar way for intermediate times~$\Tnl <
t < t_{\rm i}$.  

\begin{figure}[tb]
 \centerline{\epsfig{file=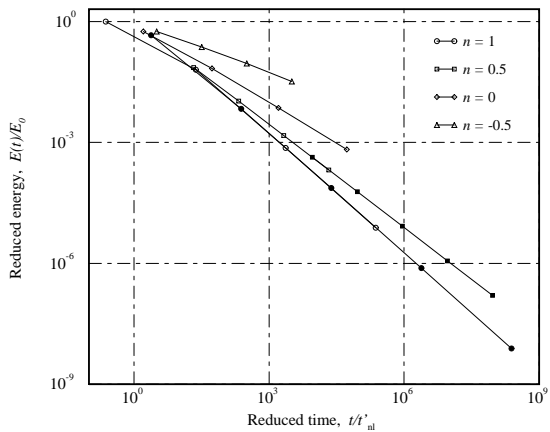,width=6cm}}
 \vspace{6mm}
 \caption{Evolution
of the computed integral energy  $E(t)$ for $n=1, 0.5, 0, -0.5$.}
 \label{figeh-.5g}
\end{figure}
 
On Fig.~\ref{figeh-.5g} the evolution of the energy $E(t)$ for 
different $n=1, 0.5, 0, -0.5$ is plotted.
The law of decay is in good agreement with the theoretical 
prediction (\ref{Etfnew}) for all time.  The constants~$a_n$ have been
measured and are shown in Table~\ref{t:n-constants}.

\begin{table}[t] 
\begin{tabular}{|r|r|r|r|r|r|r|r|r|}\hline
$n$ & -2.5 & -2.0 & -1.5 & -1.0 & -0.5 & 0.0 & 0.5 & 1.0 \\ \hline
$\gamma_n$ & $-$ & 0.94 & 0.97 & 1.09 & 1.26 & 1.43 & 1.62 & 1.93 \\
$a_n$ & $-$ & $-$ & $-$ & $-$ & 0.43 & 0.32 & 0.32 & 0.37 \\ \hline
\end{tabular}
\caption{Measured values of the universal constants~$\gamma_n$ and~$a_n$
characterizing the~$k^{-2}$ spectrum shock tail and law of energy decay
respectively.  Values marked~$-$ don't exist or couldn't be measured.}
\label{t:n-constants}
\end{table}

\section{Non-homogeneous velocity and non-homogeneous potential potential; $(-3<n<-1)$}
\label{s:nonpsi}

For the self-similar initial spectrum with $-3<n<-1$ we have a divergence of
the potential and the velocity in the infrared part of the spectrum, and of the
gradient in the ultraviolet part. Assuming that we have a cutoff wavenumber at
small scale~$\Ki$, we can still use the definition of the nonlinear time
through the gradient of velocity $\Tnl =1/\sigma_q$ (\ref{deftnLnew}), which is
equal to the the characteristic time of shock formation $t_{\rm s}$.  Due to
the divergence of the energy in the infrared part of the spectrum, the
dissipation in the shocks doesn't lead to a finite value of the energy at any
time, if there is no cutoff~$\Ki$.  Nevertheless we can still introduce the
integral scale of turbulence $L(t)$ showing the region when initial power law
spectrum $E(k,t) \sim |k|^n$ transforms to the universal spectrum $E(k,t) \sim
k^{-2}$ .  For the spectral form of principle of ``permanence of large
eddies'', we still have that integral scale grows according (\ref{Scatoffss}).
From equation~(\ref{Elargek}), we see that the amplitude of small scale part of
the spectrum decreases for $n>-2$ and increases with time for $-3<n<-2$.  

\begin{figure}[tb]
 \centerline{\epsfig{file=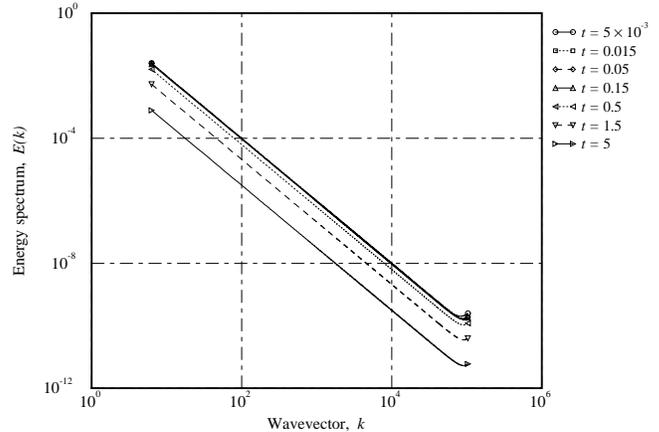,width=6cm}}
 \vspace{6mm}
 \caption{Evolution of the energy spectrum with an initial spectrum
proportional to $k^{-2}$.  One can see that the spectrum doesn't change at
all until the non-linear time of the smallest wavenumber component is reached
(near~$t\approx 0.15$).}
 \label{figh.5n}
\end{figure}

Let as start for the special case of initial spectrum with the critical index
$n=-2$ when $L(t) \sim (\alpha t)^2$. From equation (\ref{ssspectr}), we see
that the spectrum does not change in time. In Fig.~\ref {figh.5n}, energy
spectra $E(k,t)$ are shown at  different moments of time from $t/\Tnl = 0.005$
to $t/\Tnl=5$.  And we really see that it is only when $t/\Tnl\gg 1$ that the
spectrum begins to change, simply decaying in amplitude without changing its
shape.  But even if it is not apparent in the spectrum, we have an evolution in
time of each realization and of other statistical properties, like shock
probability distribution or higher moments of the velocity.

\begin{figure}[tb]
 \centerline{\epsfig{file=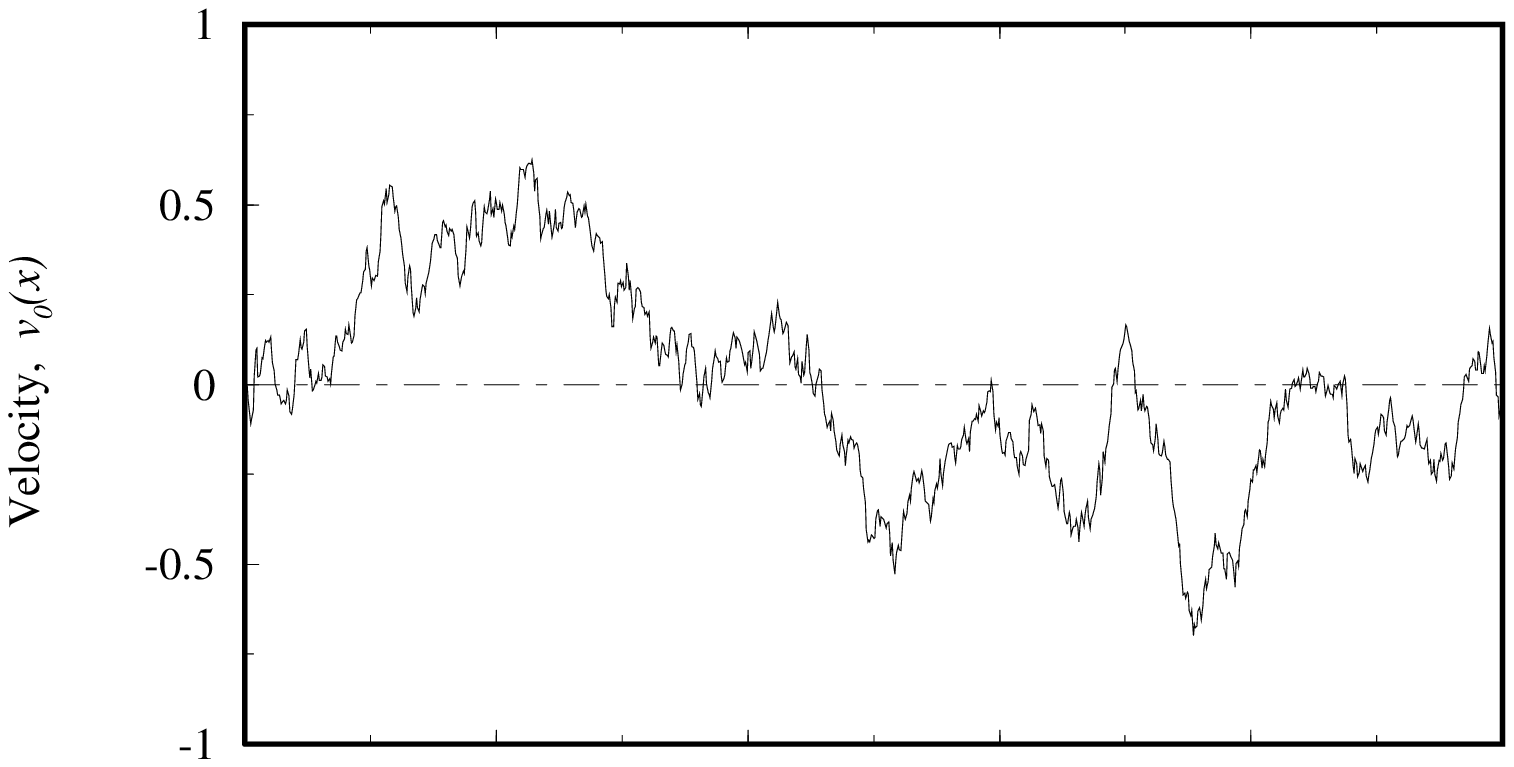,width=6cm}}
 \vspace{2mm}
 \centerline{\epsfig{file=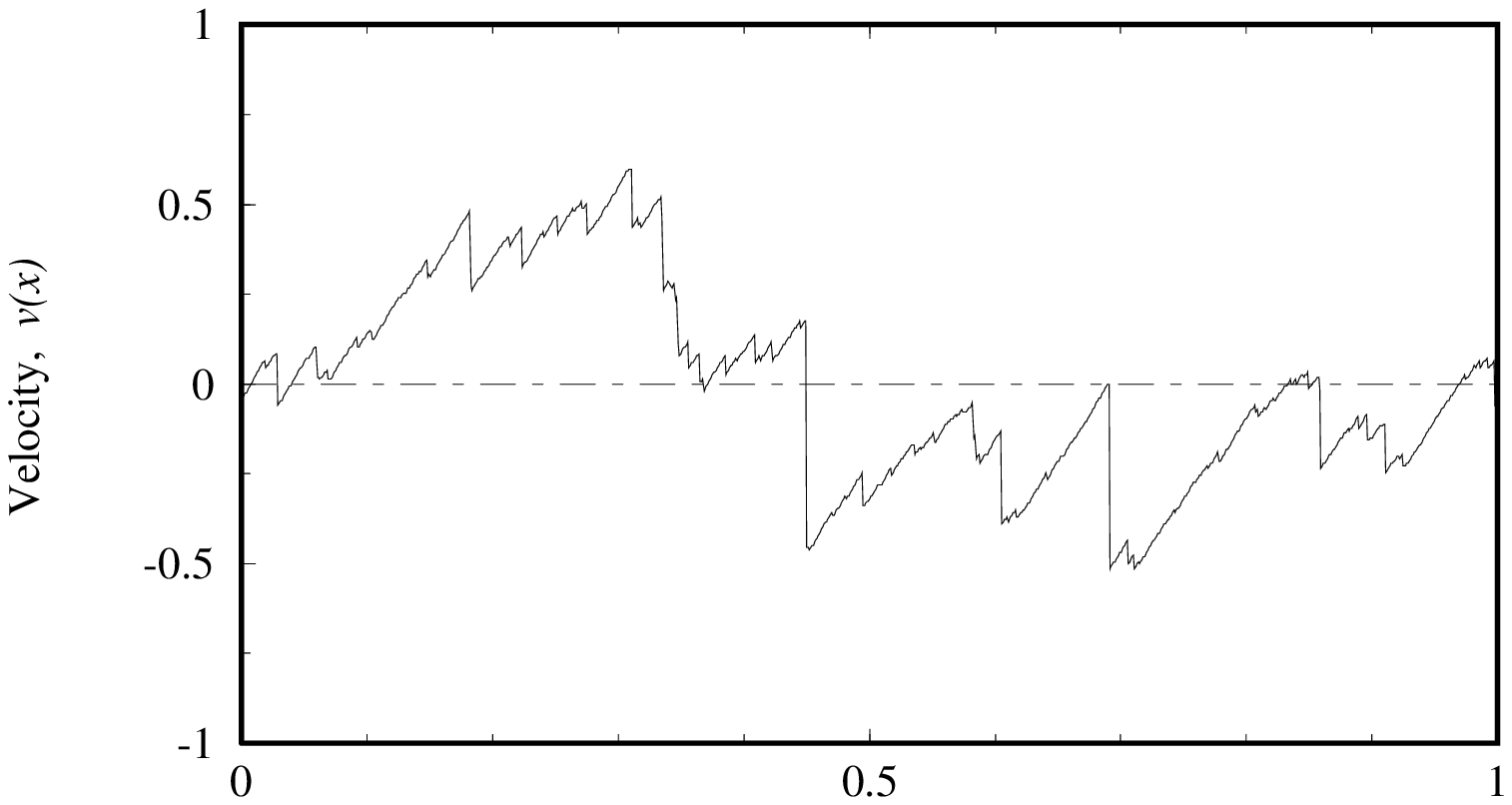,width=6cm}}
 \vspace{6mm}
 \caption{Evolution of the velocity field for one particular realization
with an initial spectrum~$k^{-2}$.  Upper plot is the Gaussian Brownian
motion initial condition at~$t=0$ and lower plot is the velocity field
at~$t=0.15$, just before the decay of the smallest wavenumber component.
Both signals have the same power spectrum.
Note that the largest scales are non-linearly distorted, but their amplitude
is globally preserved, illustrating the principle of persistence of large
eddies.}
 \label{fighreal.5n}
\end{figure}

\begin{figure}[tb]
 \centerline{\epsfig{file=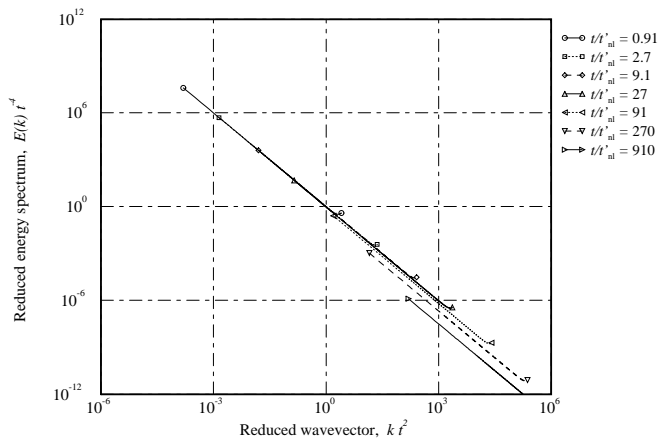,width=6cm}}
 \vspace{6mm}
 \caption{Reduced energy spectra $E(k,t)\,t^{-4}$ as a function of the
reduced wavenumber~$k\,L(t)=k\,t^2$ for an initial spectrum~$k^{-2}$.
Note that the spectrum rescaling works even if this case with a factor
{\em decreasing\/} with time, and allows the separation of the~$k^{-2}$
spectra in differents zones coming from the initial conditions ($k\,L(t)<1$)
of from the shocks ($k\,L(t)>1$).}
 \label{figh.5r}
\end{figure}

On Fig.~\ref{fighreal.5n}, the evolution of a realization of a velocity field
with a~$k^{-2}$ spectrum is plotted at different times.  It is easy to see,
that even if the spectrum does not change, the characteristic distance between
the shocks which is proportional to the scale $L(t)$ increases with time.  The
signal switches continuously from a Gaussian Brownian motion with random phases
to a triangular wave with aligned phases, but these signals have the same power
spectrum, and the amplitude of the spectrum is preserved by the Burgers
evolution.  The evolution of the spectrum in numerical simulations starts and
the energy begins to decay at the time $t_{\rm nl,1}$ when the integral scale
of turbulence $L(t)$ reaches the size of the box $L_{\rm box}$.  This can also
be seen in~Fig.~\ref{figh.5r} showing the spectra rescaled with time for
Gaussian Brownian motion initial conditions.  One can see that the
initial~$k^{-2}$ spectrum visible for~$k\,L(t)<1$ is continuated with
the~$k^{-2}$ spectrum of the shocks at late times for~$k\,L(t)>1$, but that the
rescaling allows the separation of both.  For a given finite-size realization
of the signal with both lower and upper wavenumber cutoffs~$\Ki$ and~$\Ku$, the
spectrum will slide along the curve in~Fig.~\ref{figh.5r} until~$\Ki\,L(t) >
1$, that is the rescaled wavenumbers~$k\,L(t)>1$ are larger than~1 for
all~$k$s, at which time the spectrum will begin to decay linearly (as shown
in~Fig.~\ref{figh.5r} for~$t/\Tnl > 27$).

\begin{figure}[tb]
 \centerline{\epsfig{file=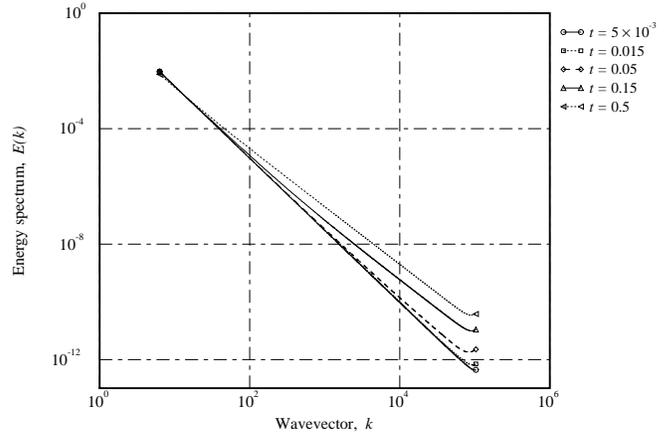,width=6cm}}
 \vspace{6mm}
 \caption{Evolution of the energy spectrum with an initial spectrum
proportional to $|k|^{-2.5}$. One can see that the amplitude of the
high-frequency components {\em increase\/} with time, the $|k|^{-2.5}$
spectrum transforming to a~$k^{-2}$ tail, whose energy really comes from
the smallest wavenumber components.  These have a very slightly decreasing
amplitude, so that the global energy is nearly constant, but slightly
decays in time.}
 \label{figh.75n}
\end{figure}

In fig.~\ref{figh.75n} the energy spectra $E(k,t)$ with an initial power law
spectrum $E(k,t) \sim |k|^n$ with $n=-2.5$ is shown at  different moments of
time from $t/\Tnl = 6.9 \times 10^{-2}$ to $t/\Tnl=6.9$.  We see the generation
of universal $k^{-2}$ tail which amplitude increases with time and switching
point between $k^{-2.5}$ and $k^{-2}$ parts of the spectrum moved to the small
wavenumbers. The value of dimensionless constant~$\gamma_n$ in the
dimensionless spectrum in (\ref{Ekofn}) are shown in Table~\ref{t:n-constants}.
Note that the constant for~$n=-2.5$ has not been measured as the energy begins
to decay immediately as soon as the~$k^{-2}$ tail is established.  This effect
also perturbs the measurement of~$\gamma_n$ for~$n<-1$, because in a
finite-size system, some (small) dissipation occurs as soon as shocks are
present, and so the amplitude of the~$k^{-2}$ tail is diminished.  This
explains why we measure for instance an amplitude~$\gamma_{-2}=0.94$ at large
times, even if~$\gamma_{-2}=1.0$ should be observed.

\begin{figure}[tb]
 \centerline{\epsfig{file=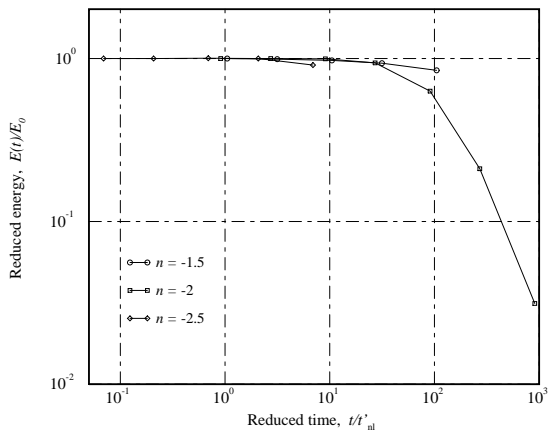,width=6cm}}
 \vspace{6mm}
 \caption{Evolution
of the computed integral energy $E(t)$ for $n=-1.5, -2, 0, -2.5$ as a 
function of $t/\Tnl$.}
 \label{figeh.5g}
\end{figure}

We can also introduce the other nonlinear time is a time of nonlinear decay of
first harmonic $t_{\rm nl,1}=k_1/A_1$. For~$-3<n<1$, we have no significant
decay of the energy of turbulence until $t \geq t_{\rm nl,1}$.  The constancy
of the energy until much later than~$\Tnl$ is also evidenced in
Fig.~\ref{figeh.5g} for $n=-1.5, -2, -2.5$.  As the energy doesn't decay
until~$t_{\rm nl,1}$, the ``constants''~$a_n$ don't exist for~$n<-1$ and are
thus not shown in Table~\ref{t:n-constants}.

\section{Discussion}
\label{s:discussion}

In this work we have reconsidered again the classical problem of the spectral
properties of solutions of Burgers' equation for long times, when the initial
velocity and velocity potential are stationary Gaussian processes.  We have
shown in greater detail how the self-similar (and not self-similar) regimes are
realized with initial conditions that are only self-similar over a finite
range.  The range in which self-similarity can be observed (or not observed)
changes in wave-number space with time, in a way that depends both on the
initial spectral slope, and on the low-$k$ and high-$k$ cutoffs in the initial
data.

Depending on the statistical properties of initial velocity and potential on
can introduce the following regions on $n$-axis. They are: homogeneous velocity
and homogeneous potential: $(n>1)$ with subinterval $(1<n<2)$; homogeneous
velocity and and non-homogeneous potential:  $(-1<n<1)$; non-homogeneous
velocity and non-homogeneous potential: $(-3<n<-1)$ with some critical point
$n=-2$.  The common properties of turbulence is the self-similar behavior,
determined by only one scale $L(t)$ - the integral scale, for a range of time
and wave numbers even in the presence of high- or low-frequency cutoffs.  But
the type of self-similarity is different in different region on ``$n$-axis''
and is determined by the  properties of initial potential.  High-resolution
numerical simulations have been performed confirming both scaling predictions
and analytical asymptotic theory.

For the case $(n>1)$ we have verified by numerical experiments the asymptotic
theory derived previously by several groups
\cite{Kida,FF,GS81,Molchanov,GSAFT97}.  The main results are\,: at very large
times the spectrum tends to a limiting shape, proportional to $k^2$ at small
wave numbers, and to $k^{-2}$ at large wave numbers, such that evolution shape
is determined by peak wave number, $k_{L}(t) \sim 1/L(t)$. Due to the merging
of the shocks the integral scale increases with time $L(t)\simeq
(t\sigma_{\psi})^{1/2}\ln^{-1/4}\left(t\sigma_{v}^2/ 2\pi
\sigma_{\psi}\right)$. So asymptotically the evolution of Burgers turbulence
and, in particular, the law of energy decay is determined only by the variance
of potential $\sigma_{\psi}^2$.
 
For large, but finite time, we have breakdown of self-similarity of the
spectrum when $1<n<2$.  The spectrum then has three scaling regions\,: first, a
$|k|^n$ region at very small $k<\Ks(t)$ with a time-independent constant,
second a $k^2$ region at intermediate wavenumbers $\Ks(t)<k<k_L(t)$ with
increasing amplitude, and, finally, the usual $k^{-2}$ region at $k>k_L(t)$.
The relative part of the spectrum with $|k|^n$ region decreases with time
$\Ks(t)/k_L(t) \sim \left(t/\tnl \right)^{- (n-1)/2(2-n)}$.

In the case of finite viscosity, if one introduces an instantaneous Reynolds
number $\hbox{Re}(t) \sim L(t) E^{1/2}(t)/\nu$ based on viscosity, the typical
velocity and typical spatial scale at time $t$, it means that $\hbox{Re}(t)\sim
\hbox{Re}_0 (\log t/\tnl)^{-1/2}$.  Within dimensional estimates, the Reynolds
number would be constant in time.  On a practical level, we have thus
established that the inviscid approximation is not valid for arbitrary long
times.  After a time, which is very long if the initial Reynolds number is
large, about $ \tnl\exp(\hbox{Re}_0^2)$, the viscous term in (\ref{BE}) becomes
comparable to the inertial term everywhere, and can not be neglected.  

When~$n$ is less than one, the large-scale part of the spectrum $E(k)=\alpha^2
|k|^n$ is preserved in time and the global evolution is self-similar, so that
scaling arguments perfectly predict the behavior in time of the integral scale
$L(t)= (\alpha t)^{2/(3+n)}$.  For $-1<n<1$, the energy also decay as a power
law $E(t) \sim \alpha ^{4/(3+n)} t^{-2(n+1)/(3+n)}$.  In case of finite
viscosity the increasing of the integral scale is faster the decay of the
energy and we have for the Reynolds number $\hbox{Re}(t) \sim t^{(1-n)/(3+n)}$,
i.e. the Reynolds  number increases  with time and the shape of the wave
becomes more and more nonlinear.  This last point is true only in the case when
we have not cutoff wavenumber at large scale.  In numerical simulations in a
finite box, the final behavior will always be the linearly decaying sinusoidal
wave with period equal to the size of the box.

\section {Acknowledgments.} 

We have benefited from discussions with U.~Frisch and A.~Saichev. This work was
supported by the French Ministry of Higher Education, RFBR- 02-02-17374,
LSS-838.2003.2 and ``Universities of Russia'' grants, and by the Swedish
Research Council and the Swedish Institute.

\section{Appendix. Numerical work}

\label{s:numerical}.

\subsection{Normalizations of initial spectrum}
\label{ss:normalizations}
We use the following smooth cutoff of the initial power spectrum\,:
\begin{equation}
E_0(k) = \alpha^2 |k|^n
\mbox{\Large e}^{-\frac{k^2}{2k_0^2}}.
\label{NIS}
\end{equation}

The variances of the velocity, the velocity  gradient and potential
can  be computed to be:
\begin{eqnarray}
\label{Nv}
\sigma_v^2=\langle v_0^2 \rangle & = & \alpha^2 
\Gamma\left(\frac{n+1}{2} \right) k_0^{n+1},  \\
\label{Nq}
\sigma_q^2=\langle q_0^2 \rangle & = & \alpha^2 
\Gamma\left(\frac{n+3}{2} \right) k_0^{n+3}, \\
\label{Npsi}
\sigma_{\psi}^2=\langle \psi_0^2 \rangle & = & \alpha^2 
\Gamma\left(\frac{n-1}{2} \right) k_0^{n-1}, 
\end{eqnarray}
where $\Gamma(x)$ is the gamma-function. 

 From this equation it is easy to see the critical points of $n$
connected with the divergence at small wave numbers.
\subsection{Generation of initial conditions}
\label{ss:generation}

The scale of the box in the numerical experiment is taken as unit of space,
so that wavenumbers~$k$ go from $2\pi$ to $N\pi$, where~$N$ is the number of points used in the simulation, typically ranging from~$N=2^{15}\simeq 3\times10^4$
to $N=2^{20} \simeq 10^6$ in our simulations.  The amplitude of the 
spectrum (\ref{NIS}) was simply taken as $\alpha=1$.
Fourier components of a Gaussian process are independent Gaussian 
variables. We therefore synthesize the initial 
potential of the velocity by first 
generating random Fourier components $a_k$ distributed according to 
\[
p(a_k) = \frac{1}{\sqrt{2\pi\sigma_k^2} } \exp \left(-\frac{a_k^2}{2\sigma_k^2} \right)
\]
where
\[
\sigma_k^2 = E_{\psi}(k) dk.
\]
And we use the well known relation between the power spectra of the
process and of its integral
\begin{equation}
E_{\psi}(k) = k^{-2} E_0(k)
\label{SPV}
\end{equation}
where the form of $E_0(k)$ is chosen with a smooth cutoff at large $k$ 
according to (\ref{NIS}).

By inverse Fourier transforming the components, we obtain the initial potential
in real space, from which we can obtain the potential at any time using the
Legendre Transform~(\ref{MAX}) (see~\cite{Noullez,Vergassola}).
Repeating the whole process many times with different realizations
of~$\psi_0(x)$, we sample the desired ensemble of Gaussian initial conditions.  

\subsection{Fast Legendre Transforms}
\label{ss:flg}

In numerical simulations the initial data are always generated as  
a discrete set of $N$ points. It could be assumed naively that 
the number of operations necessary to compute the maximization 
(\ref{MAX}) for all values of $x$ scales as $O(N^2)$. It may however be shown, 
using (\ref{MAX}) that $a(x)$ is a nondecreasing function of $x$.
The number of operations needed in an ordered search therefore scales 
as $O(N\log_2 N)$ when using the so-called Fast Legendre Transform
procedure \cite{SheAurellFrisch,Noullez,Vergassola}.

\bibliographystyle{apsrev}
\bibliography{decay}

\begin{thebibliography}{33}
\expandafter\ifx\csname natexlab\endcsname\relax\def\natexlab#1{#1}\fi
\expandafter\ifx\csname bibnamefont\endcsname\relax
  \def\bibnamefont#1{#1}\fi
\expandafter\ifx\csname bibfnamefont\endcsname\relax
  \def\bibfnamefont#1{#1}\fi
\expandafter\ifx\csname citenamefont\endcsname\relax
  \def\citenamefont#1{#1}\fi
\expandafter\ifx\csname url\endcsname\relax
  \def\url#1{\texttt{#1}}\fi
\expandafter\ifx\csname urlprefix\endcsname\relax\def\urlprefix{URL }\fi
\providecommand{\bibinfo}[2]{#2}
\providecommand{\eprint}[2][]{\url{#2}}

\bibitem[{\citenamefont{Burgers}(1974)}]{Burgers}
\bibinfo{author}{\bibfnamefont{J.}~\bibnamefont{Burgers}},
  \emph{\bibinfo{title}{The Nonlinear Diffusion Equation}}
  (\bibinfo{publisher}{D. Reidel, Publ. Co.}, \bibinfo{year}{1974}).

\bibitem[{\citenamefont{Whitham}(1974)}]{Whitham}
\bibinfo{author}{\bibfnamefont{G.}~\bibnamefont{Whitham}},
  \emph{\bibinfo{title}{Linear and Nonlinear Waves}}
  (\bibinfo{publisher}{Wiley}, \bibinfo{year}{1974}).

\bibitem[{\citenamefont{Rudenko and Soluyan}(1977)}]{RudenkoSoluyan}
\bibinfo{author}{\bibfnamefont{O.}~\bibnamefont{Rudenko}} \bibnamefont{and}
  \bibinfo{author}{\bibfnamefont{S.}~\bibnamefont{Soluyan}},
  \emph{\bibinfo{title}{Theoretical foundations of nonlinear acoustics}}
  (\bibinfo{publisher}{New York: Plenum Press}, \bibinfo{year}{1977}).

\bibitem[{\citenamefont{Kuramoto}(1985)}]{Kuramoto}
\bibinfo{author}{\bibfnamefont{Y.}~\bibnamefont{Kuramoto}},
  \emph{\bibinfo{title}{Chemical Oscillations, Waves and Turbulence}}
  (\bibinfo{publisher}{Berlin: Springer Verlag}, \bibinfo{year}{1985}).

\bibitem[{\citenamefont{Gurbatov et~al.}(1991)\citenamefont{Gurbatov, Malakhov,
  and Saichev}}]{Gurbatov}
\bibinfo{author}{\bibfnamefont{S.}~\bibnamefont{Gurbatov}},
  \bibinfo{author}{\bibfnamefont{A.}~\bibnamefont{Malakhov}}, \bibnamefont{and}
  \bibinfo{author}{\bibfnamefont{A.}~\bibnamefont{Saichev}},
  \emph{\bibinfo{title}{Nonlinear random waves and turbulence in nondispersive
  media: waves, rays, particles}} (\bibinfo{publisher}{Manchester University
  Press}, \bibinfo{year}{1991}).

\bibitem[{\citenamefont{Woyczynski}(1998)}]{WW98}
\bibinfo{author}{\bibfnamefont{W.}~\bibnamefont{Woyczynski}},
  \emph{\bibinfo{title}{Burgers--KPZ Turbulence. Gottingen Lectures.}}
  (\bibinfo{publisher}{Springer-Verlag}, \bibinfo{year}{1998}).

\bibitem[{\citenamefont{Gurbatov et~al.}(1989)\citenamefont{Gurbatov, Saichev,
  and Shandarin}}]{GurbatovSaichev}
\bibinfo{author}{\bibfnamefont{S.}~\bibnamefont{Gurbatov}},
  \bibinfo{author}{\bibfnamefont{A.}~\bibnamefont{Saichev}}, \bibnamefont{and}
  \bibinfo{author}{\bibfnamefont{S.}~\bibnamefont{Shandarin}},
  \bibinfo{journal}{Mon.Not.R.Astr.Soc.} \textbf{\bibinfo{volume}{236}},
  \bibinfo{pages}{385} (\bibinfo{year}{1989}).

\bibitem[{\citenamefont{Shandarin and Zel'dovich}(1989)}]{Shandarin}
\bibinfo{author}{\bibfnamefont{S.}~\bibnamefont{Shandarin}} \bibnamefont{and}
  \bibinfo{author}{\bibfnamefont{Y.}~\bibnamefont{Zel'dovich}},
  \bibinfo{journal}{Rev. Mod. Phys.} \textbf{\bibinfo{volume}{61}},
  \bibinfo{pages}{185} (\bibinfo{year}{1989}).

\bibitem[{\citenamefont{Vergassola et~al.}(1994)\citenamefont{Vergassola,
  Dubrulle, Frisch, and Noullez}}]{Vergassola}
\bibinfo{author}{\bibfnamefont{M.}~\bibnamefont{Vergassola}},
  \bibinfo{author}{\bibfnamefont{B.}~\bibnamefont{Dubrulle}},
  \bibinfo{author}{\bibfnamefont{U.}~\bibnamefont{Frisch}}, \bibnamefont{and}
  \bibinfo{author}{\bibfnamefont{A.}~\bibnamefont{Noullez}},
  \bibinfo{journal}{Astro. Astrophys.} \textbf{\bibinfo{volume}{289}},
  \bibinfo{pages}{325} (\bibinfo{year}{1994}).

\bibitem[{\citenamefont{Sinai}(1992)}]{Sinai}
\bibinfo{author}{\bibfnamefont{Y.}~\bibnamefont{Sinai}},
  \bibinfo{journal}{Comm. Math. Phys.} \textbf{\bibinfo{volume}{148}},
  \bibinfo{pages}{601} (\bibinfo{year}{1992}).

\bibitem[{\citenamefont{She et~al.}(1992)\citenamefont{She, Aurell, and
  Frisch}}]{SheAurellFrisch}
\bibinfo{author}{\bibfnamefont{Z.}~\bibnamefont{She}},
  \bibinfo{author}{\bibfnamefont{E.}~\bibnamefont{Aurell}}, \bibnamefont{and}
  \bibinfo{author}{\bibfnamefont{U.}~\bibnamefont{Frisch}},
  \bibinfo{journal}{Comm. Math. Phys.} \textbf{\bibinfo{volume}{148}},
  \bibinfo{pages}{623} (\bibinfo{year}{1992}).

\bibitem[{\citenamefont{Schwartz and Edwards}(1992)}]{Schwartz}
\bibinfo{author}{\bibfnamefont{M.}~\bibnamefont{Schwartz}} \bibnamefont{and}
  \bibinfo{author}{\bibfnamefont{S.}~\bibnamefont{Edwards}},
  \bibinfo{journal}{Europhys. Lett.} \textbf{\bibinfo{volume}{20}},
  \bibinfo{pages}{301} (\bibinfo{year}{1992}).

\bibitem[{\citenamefont{Kardar et~al.}(1986)\citenamefont{Kardar, Parisi, and
  Zhang}}]{Kardar}
\bibinfo{author}{\bibfnamefont{M.}~\bibnamefont{Kardar}},
  \bibinfo{author}{\bibfnamefont{G.}~\bibnamefont{Parisi}}, \bibnamefont{and}
  \bibinfo{author}{\bibfnamefont{Y.}~\bibnamefont{Zhang}},
  \bibinfo{journal}{Phys. Rev. Lett.} \textbf{\bibinfo{volume}{56}},
  \bibinfo{pages}{889} (\bibinfo{year}{1986}).

\bibitem[{\citenamefont{Barab\'asi and Stanley}(1995)}]{BS95}
\bibinfo{author}{\bibfnamefont{A.-L.} \bibnamefont{Barab\'asi}}
  \bibnamefont{and} \bibinfo{author}{\bibfnamefont{H.}~\bibnamefont{Stanley}},
  \emph{\bibinfo{title}{Fractal Concepts in Surface Growth}}
  (\bibinfo{publisher}{Cambridge University Press}, \bibinfo{year}{1995}).

\bibitem[{\citenamefont{Frachebourg and Martin}(2000)}]{Fra00}
\bibinfo{author}{\bibfnamefont{L.}~\bibnamefont{Frachebourg}} \bibnamefont{and}
  \bibinfo{author}{\bibfnamefont{P.~A.} \bibnamefont{Martin}},
  \bibinfo{journal}{J. Fluid Mechanics} \textbf{\bibinfo{volume}{417}},
  \bibinfo{pages}{323} (\bibinfo{year}{2000}).

\bibitem[{\citenamefont{Molchan}(1997)}]{Molchan}
\bibinfo{author}{\bibfnamefont{G.}~\bibnamefont{Molchan}}, \bibinfo{journal}{J.
  Stat. Phys.} \textbf{\bibinfo{volume}{88}}, \bibinfo{pages}{1139}
  (\bibinfo{year}{1997}).

\bibitem[{\citenamefont{Frisch}(1995)}]{Frisch}
\bibinfo{author}{\bibfnamefont{U.}~\bibnamefont{Frisch}},
  \emph{\bibinfo{title}{Turbulence: the Legacy of A.N.~Kolmogorov}}
  (\bibinfo{publisher}{Cambridge University Press}, \bibinfo{year}{1995}).

\bibitem[{\citenamefont{Gurbatov et~al.}(1997)\citenamefont{Gurbatov,
  Simdyankin, Aurell, Frisch, and Toth}}]{GSAFT97}
\bibinfo{author}{\bibfnamefont{S.}~\bibnamefont{Gurbatov}},
  \bibinfo{author}{\bibfnamefont{S.}~\bibnamefont{Simdyankin}},
  \bibinfo{author}{\bibfnamefont{E.}~\bibnamefont{Aurell}},
  \bibinfo{author}{\bibfnamefont{U.}~\bibnamefont{Frisch}}, \bibnamefont{and}
  \bibinfo{author}{\bibfnamefont{G.}~\bibnamefont{Toth}}, \bibinfo{journal}{J.
  Fluid Mech} \textbf{\bibinfo{volume}{344}}, \bibinfo{pages}{349}
  (\bibinfo{year}{1997}).

\bibitem[{\citenamefont{Kolmogorov}(1941)}]{Kolmogorov1941b}
\bibinfo{author}{\bibfnamefont{A.}~\bibnamefont{Kolmogorov}},
  \bibinfo{journal}{Dokl. Akad. Nauk SSSR} \textbf{\bibinfo{volume}{31}},
  \bibinfo{pages}{538} (\bibinfo{year}{1941}).

\bibitem[{\citenamefont{Loitsyansky}(1939)}]{Loitsyanski}
\bibinfo{author}{\bibfnamefont{L.}~\bibnamefont{Loitsyansky}},
  \bibinfo{journal}{Trudy Tsentr. Aero.-Gidrodin. Inst} pp.
  \bibinfo{pages}{3--23} (\bibinfo{year}{1939}).

\bibitem[{\citenamefont{Eyink and Thomson}(2000)}]{Eyink00}
\bibinfo{author}{\bibfnamefont{G.}~\bibnamefont{Eyink}} \bibnamefont{and}
  \bibinfo{author}{\bibfnamefont{D.}~\bibnamefont{Thomson}},
  \bibinfo{journal}{Physics of Fluids} \textbf{\bibinfo{volume}{12}},
  \bibinfo{pages}{477} (\bibinfo{year}{2000}).

\bibitem[{\citenamefont{Hopf}(1950)}]{Hopf}
\bibinfo{author}{\bibfnamefont{E.}~\bibnamefont{Hopf}}, \bibinfo{journal}{Comm.
  Pure Appl. Mech.} \textbf{\bibinfo{volume}{3}}, \bibinfo{pages}{201}
  (\bibinfo{year}{1950}).

\bibitem[{\citenamefont{Cole}(1951)}]{Cole}
\bibinfo{author}{\bibfnamefont{J.}~\bibnamefont{Cole}},
  \bibinfo{journal}{Quart. Appl. Math.} \textbf{\bibinfo{volume}{9}},
  \bibinfo{pages}{225} (\bibinfo{year}{1951}).

\bibitem[{\citenamefont{Kida}(1979)}]{Kida}
\bibinfo{author}{\bibfnamefont{S.}~\bibnamefont{Kida}}, \bibinfo{journal}{J.
  Fluid Mech.} \textbf{\bibinfo{volume}{93}}, \bibinfo{pages}{337}
  (\bibinfo{year}{1979}).

\bibitem[{\citenamefont{Gurbatov and Saichev}(1981)}]{GS81}
\bibinfo{author}{\bibfnamefont{S.}~\bibnamefont{Gurbatov}} \bibnamefont{and}
  \bibinfo{author}{\bibfnamefont{A.}~\bibnamefont{Saichev}},
  \bibinfo{journal}{Sov. Phys. JETP} \textbf{\bibinfo{volume}{53}},
  \bibinfo{pages}{347} (\bibinfo{year}{1981}).

\bibitem[{\citenamefont{Fournier and Frisch}(1983)}]{FF}
\bibinfo{author}{\bibfnamefont{J.~D.} \bibnamefont{Fournier}} \bibnamefont{and}
  \bibinfo{author}{\bibfnamefont{U.}~\bibnamefont{Frisch}},
  \bibinfo{journal}{J. de M\'ec. Th\'eor. et Appl.}
  \textbf{\bibinfo{volume}{2}}, \bibinfo{pages}{699} (\bibinfo{year}{1983}).

\bibitem[{\citenamefont{Molchanov et~al.}(1995)\citenamefont{Molchanov,
  Surgailis, and Woyczynski}}]{Molchanov}
\bibinfo{author}{\bibfnamefont{S.}~\bibnamefont{Molchanov}},
  \bibinfo{author}{\bibfnamefont{D.}~\bibnamefont{Surgailis}},
  \bibnamefont{and}
  \bibinfo{author}{\bibfnamefont{W.}~\bibnamefont{Woyczynski}},
  \bibinfo{journal}{Comm. Math. Phys.} \textbf{\bibinfo{volume}{168}},
  \bibinfo{pages}{209} (\bibinfo{year}{1995}).

\bibitem[{\citenamefont{Leadbetter et~al.}(1983)\citenamefont{Leadbetter,
  Lindgren, and Rootzen}}]{LeadbetterLindgrenRootzen}
\bibinfo{author}{\bibfnamefont{M.}~\bibnamefont{Leadbetter}},
  \bibinfo{author}{\bibfnamefont{G.}~\bibnamefont{Lindgren}}, \bibnamefont{and}
  \bibinfo{author}{\bibfnamefont{H.}~\bibnamefont{Rootzen}},
  \emph{\bibinfo{title}{Extremes and Related Properties of Random Sequences and
  Processes}} (\bibinfo{publisher}{Springer, Berlin}, \bibinfo{year}{1983}).

\bibitem[{\citenamefont{Gurbatov and Malakhov}(1977)}]{GurbatovMalakhov1977}
\bibinfo{author}{\bibfnamefont{S.}~\bibnamefont{Gurbatov}} \bibnamefont{and}
  \bibinfo{author}{\bibfnamefont{A.}~\bibnamefont{Malakhov}},
  \bibinfo{journal}{Sov. Phys. Acoust.} \textbf{\bibinfo{volume}{23}},
  \bibinfo{pages}{325} (\bibinfo{year}{1977}).

\bibitem[{\citenamefont{Avellaneda and E}(1995)}]{AvellanedaE}
\bibinfo{author}{\bibfnamefont{M.}~\bibnamefont{Avellaneda}} \bibnamefont{and}
  \bibinfo{author}{\bibfnamefont{W.}~\bibnamefont{E}}, \bibinfo{journal}{Comm.
  Math. Phys.} \textbf{\bibinfo{volume}{172}}, \bibinfo{pages}{13}
  (\bibinfo{year}{1995}).

\bibitem[{\citenamefont{Aurell et~al.}(1993)\citenamefont{Aurell, Gurbatov, and
  Wertgeim}}]{Aurell}
\bibinfo{author}{\bibfnamefont{E.}~\bibnamefont{Aurell}},
  \bibinfo{author}{\bibfnamefont{S.}~\bibnamefont{Gurbatov}}, \bibnamefont{and}
  \bibinfo{author}{\bibfnamefont{I.}~\bibnamefont{Wertgeim}},
  \bibinfo{journal}{Phys.Letters A} \textbf{\bibinfo{volume}{182}},
  \bibinfo{pages}{109} (\bibinfo{year}{1993}).

\bibitem[{\citenamefont{Gurbatov and Pasmanik}(1999)}]{GP99}
\bibinfo{author}{\bibfnamefont{S.}~\bibnamefont{Gurbatov}} \bibnamefont{and}
  \bibinfo{author}{\bibfnamefont{G.}~\bibnamefont{Pasmanik}},
  \bibinfo{journal}{J. Experimental Theoretical Physics}
  \textbf{\bibinfo{volume}{88}}, \bibinfo{pages}{309} (\bibinfo{year}{1999}).

\bibitem[{\citenamefont{Noullez and Vergassola}(1994)}]{Noullez}
\bibinfo{author}{\bibfnamefont{A.}~\bibnamefont{Noullez}} \bibnamefont{and}
  \bibinfo{author}{\bibfnamefont{M.}~\bibnamefont{Vergassola}},
  \bibinfo{journal}{J. Sci. Comp.} \textbf{\bibinfo{volume}{9}},
  \bibinfo{pages}{259} (\bibinfo{year}{1994}).

\end{thebibliography}

\end{document}